\documentclass[prd,showpacs,nofootinbib,preprintnumbers]{revtex4}

\usepackage{graphicx}
\usepackage{feynmf}
\usepackage{amssymb}
\usepackage{pifont}

\def\bfnabla{\mbox{\boldmath $\nabla$}}
\def\bfsigma{\mbox{\boldmath $\sigma$}}

\newcommand{\be}{
\begin{equation}
}{\bf }
\newcommand{\ee}{
\end{equation}
}
\newcommand{\bea}{
\begin{eqnarray}
}
\newcommand{\eea}{
\end{eqnarray}
}
\newcommand{\sla}{\!\!\!\!/ \,}

\def\siml{{\ \lower-1.2pt\vbox{\hbox{\rlap{$<$}\lower6pt\vbox{\hbox{$\sim$}}}}\ }}

\begin{document}

\title{Non-relativistic bound states at finite temperature (II): the muonic hydrogen}
\author{Miguel Angel Escobedo}
\author{Joan Soto}
\affiliation{Departament d'Estructura i Constituents de la Mat\`eria and Institut de Ci\`encies del Cosmos, Universitat de Barcelona\\
Diagonal 647, E-08028 Barcelona, Catalonia, Spain}
\preprint{UB-ECM-PF 09/15 $ , \;$ ICCUB-10-105}

\begin{abstract}

We illustrate how to apply modern effective field theory techniques and dimensional regularization to factorise the various scales which appear in QED bound states at finite temperature. We focus here on the muonic hydrogen atom. Vacuum polarization effects make the physics of this atom at finite temperature very close to that of heavy quarkonium states. We comment on the implications of our results for these states in the quark gluon plasma.  
In particular, we estimate the effects of a finite charm quark mass in the dissociation temperature of bottomonium.

\end{abstract}

\maketitle

\section{Introduction}

In a previous paper \cite{Escobedo:2008sy} we showed how to apply modern effective field theory (EFT) techniques to the hydrogen atom at finite temperature. They provide a systematic way to separate the physics occurring at the various dynamical scales
involved in that system, which makes calculations simple and transparent. The main motivation of that work was to pave the way to a QCD based quantitative study of heavy quarkonium states in the quark gluon plasma (several works in this direction have recently appeared in the literature \cite{Laine:2006ns,Laine:2007gj,Burnier:2007qm,Beraudo:2007ky,Brambilla:2008cx,Burnier:2008ia}), which share with the hydrogen atom a number of important features. The main qualitative difference, as far as the bound state dynamics is concerned, between heavy quarkonium states and the hydrogen atom is that vacuum polarization effects in the latter are very suppressed. For muonic hydrogen, however, the vacuum polarization effects 
provide the leading corrections to the Coulomb potential, as it is the case for heavy quarkonium states. This is our main motivation to study muonic hydrogen in detail here.  

Muonic hydrogen is under current research at the Paul Scherrer Institute in the so called Muonic Hydrogen Lamb Shift experiment \cite{PSI}. It allows for precision tests of QED which, among other things, probe the electromagnetic structure of the proton \cite{DiGiacomo:1969tj} or the size of the proton \cite{PSI2}. Theoretical calculations, on one hand, have achieved an impressive precision \cite{Ivanov:2009aa,Borie:2004fv,Kinoshita:1998jg,Kinoshita:1998jf,Pachucki:1996zza,Pineda:2002as}, and a number of experimental results are available \cite{Klines,Fujiwara:2000yf}. However thermal effects on this atom 
due to black body radiation, or to electron-positron plasmas, have not been considered to our knowledge, either theoretically or experimentally. Current experimental facilities may now produce electron-positron plasmas \cite{Shen:2002zza}, which are also the target of intense theoretical studies \cite{Aksenov:2009dy,Aksenov} (see \cite{Thoma:2008my} for a recent review). Making muonic hydrogen atoms slowly travel through an electron-positron plasma would be an ideal experiment to probe how well we understand thermal effects in non-relativistic bound states at relatively high temperatures. Recall that an analogous experiment at lower temperature with blackbody radiation on Rydberg atoms \cite{hollberg} first detected thermal level shifts, 
 in the early eighties. This would mean a further step in taking advantage of the similarities of the electron-positron plasma with the quark-gluon plasma (see \cite{Blaizot:2001nr,Rischke:2003mt} for reviews) in order to learn about the non-trivial properties of the latter, as it has already been advocated by some authors \cite{Thoma:2008gh}.

In the center of mass frame, the proton of a muonic hydrogen is essentially at rest and the muon moves at small velocities $v\lesssim \alpha \ll 1$. Hence, the relevant scales at zero temperature are those of a non-relativistic system \cite{Caswell:1985ui}: the muon mass $m_\mu$ (hard), the typical momentum $p$ which is of the order $m_\mu\alpha /n$ (soft) and the energy $E\sim m_\mu\alpha^2 /n^2$ (ultrasoft), $n$ being the principal quantum number. Unlike the hydrogen atom case, vacuum polarization effects introduce a new scale in the muonic hydrogen atom, the electron mass $m_e$, which is of the order of the soft scale for the lower lying states ($n=1,2$), but larger for the remaining ones. At finite temperature further scales are introduced, not only $T$, the temperature, but also $eT\sim m_D$, the Debye mass, and others that will be discussed later. In order to efficiently deal with the physics at each of these scales we will use the effective theories of non-relativistic QED (NRQED) \cite{Caswell:1985ui}, suitable for energies much smaller than the hard scale, potential NRQED (pNRQED) \cite{Pineda:1997bj}, suitable for energies much smaller than the soft scale, and Hard Thermal Loop Effective Theory (HTL) \cite{Braaten:1991gm}, suitable for energies much smaller than the temperature, in a way analogous to Ref.\cite{Escobedo:2008sy}. Recall that pNRQED facilitates enormously the iteration of the Coulomb potential in ultrasoft contribution (at the scale $E$), and the HTL action does the same for soft thermal photon resummations at the scale $eT$. When the scales $eT$ and $E$ coincide the combination of pNRQED and HTL is crucial in order to obtain consistently both the iteration of the Coulomb potential and the resummation of soft thermal photons.
 
We will use the real-time formalism \cite{Lebellac}, which is mandatory for the study of the propagation of a (non-thermalised) non-relativistic system in a thermal bath. We shall restrict ourselves to temperatures much smaller than the muon mass, and hence the thermal bath does not affect the free muon propagator which remains the same as at zero temperature. The same holds true for the free proton propagator, which will be further approximated by that of a static source. Thermal propagators will in general be necessary for the photons, electrons and positrons. Recall that in the real-time formalism
a doubling of degrees of freedom is required to properly account for the thermal propagation. External propagators can only correspond to type "1" fields,
vertices contain either type "1" fields or type "2" fields. Propagators can be "11", "12", "21" or "22". When drawing Feynman diagrams we will understand that all possible types of vertices and propagators compatible with a given diagram are added up, and will not display each type explicitly (for muons and protons only the "11" propagator must be considered). The techniques and results we shall use have been reviewed in ref. \cite{Thoma:2000dc}. We reproduced the basic ones in the Appendix \ref{review}.  

We distribute the paper as follows. In Section II we study the ideal case in which the electron mass is set to zero ($m_e=0$). This not only makes calculations simpler, but also makes the system closer to the heavy quarkonium case. In section III we focus on the actual case $m_e\not= 0$. These two sections are divided in subsections in which the cases $T\ll p$, $T\sim p$ and $T\gg p$ are studied, $p$ being the typical relative momentum in the bound state (or the inverse Bohr radius). Section IV is devoted to discussion and conclusions. 

\section{$m_e=0$ case}

Let us first consider an ideal case in which the electron mass $m_e$ is taken to be zero. This case is in fact closer to the one in heavy quarkonium states, particularly in charmonium, than the actual case with $m_e\not=0$, which we will study in the next section. It has already been discussed in the past in order to clarify subtle issues on the renormalization group structure of non-relativistic effective theories \cite{Pineda:2002bv}.

\subsection{$p\gg T$}

For temperatures much smaller than the soft scale ($m_\mu \alpha$ in this case) we can study the thermal effects in the atom starting from the pNRQED Lagrangian at zero temperature, up to exponentially suppressed contributions $\sim e^{-p/T}$. This means that the potentials are the same as the ones at zero temperature. The only difference with respect to the hydrogen atom case is that these potentials contain now  ${\cal O} (\alpha )$ corrections due to vacuum polarization effects produced by virtual electron-positron pairs. Like in the hydrogen atom, the ultrasoft photons, electrons and positrons are responsible for the finite temperature effects. Our starting point in this section is then eq. (6) from \cite{Pineda:1997ie}.

\begin{eqnarray}
\label{LpNRQED}
&L_{pNRQED}=&-\int\,d^3{\bf x}{1\over 4} F_{\mu \nu}(t,{\bf x}) F^{\mu \nu}(t,{\bf x})
+\int d^3{\bf r}\,d^3{\bf R} S^{\dagger} (t,{\bf r},{\bf R})\Biggl(i\partial_0
+ {{{\bfnabla}}^2\over 2 m_\mu}+{\alpha\over \vert {\bf r}\vert}+
\nonumber
\\&&
+ {{{\bfnabla}}^4\over
8 m_\mu^3}
 +{e^2\over m_\mu^2}\left( -{c_D\over 8} +4d_2\right)\delta^3 ({\bf r})
  +ic_S{\alpha\over 4m_\mu^2}{\bfsigma} \cdot \left({{\bf r}\over \vert {\bf r}
  \vert^3}\times {\bfnabla} \right)
  \Biggr)S (t,{\bf r},{\bf R})
  \label{pnrqcd2}
  \\ &&
  +\int d^3{\bf r}\,d^3{\bf R} S^{\dagger} (t,{\bf r},{\bf R})e{\bf r} \cdot {\bf E}(t,{\bf R})S (t,{\bf r},{\bf R})+\int d^3{\bf x}\bar{e}(t,{\bf x})i\gamma^\mu D_\mu e(t,{\bf x})\nonumber
  \,.
\end{eqnarray}
where $S(t,{\bf r},{\bf R})$ is the muon wave-function field, ${\bf r}$ being its distance to the proton and ${\bf R}$ the position of the proton; $e(t,{\bf x})$ the electron Dirac field.  $\alpha=e^2/4\pi$ is the electromagnetic coupling constant, and $c_0$, $c_s$ and $d_2$ are matchings coefficients which can be found at one loop in \cite{Manohar:1997qy}.

Let us separate the cases $T\lesssim E$  and $T \gg E$, which are analysed the two following sections:

\subsubsection{$T\lesssim E$} 

\begin{figure}
\includegraphics{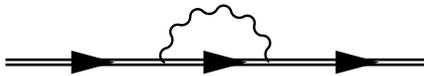}
\caption{Muonic hydrogen atom self-energy at leading order (due to ultrasoft photons). The double line stands for the propagator of the atom and the wavy line for the thermal propagator of the transverse photons. The vertex corresponds to the dipole term in the multipole expansion, see (\ref{LpNRQED}).}
\label{pNRQEDtreelevel}
\end{figure}

In this case, the leading temperature-dependent contributions are given by the diagram in fig. \ref{pNRQEDtreelevel}, in an analogous way to the hydrogen atom case. Virtual ultrasoft electron-positron pairs give rise to ${\cal O} (\alpha )$ corrections and no soft thermal photon resummation is necessary at the scale $E$.
Hence,
there is no qualitative difference with respect to the hydrogen atom, and we will not further discuss it. We refer to \cite{Escobedo:2008sy} for the relevant formulas for the spectrum and decay widths \footnote{In the case $E\gg T$ the formulas presented in \cite{Escobedo:2008sy} provide the leading contribution for $T \gg \alpha E$ only. For $T \lesssim \alpha E$ additional contributions exist \cite{Jentschura}}.

\subsubsection{$T \gg E$} 
\begin{figure}
\includegraphics{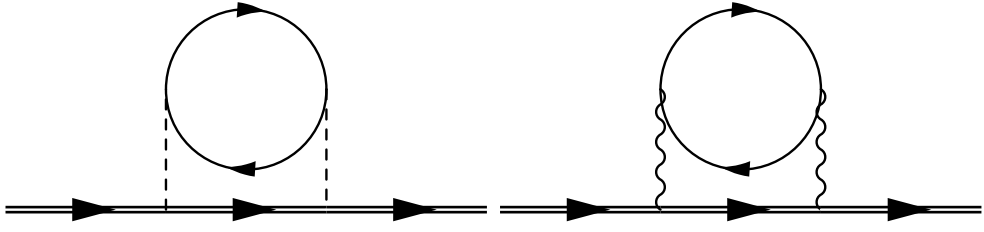}
\caption{Muonic hydrogen atom self-energy at next-to-leading order (due to ultrasoft photons, electrons and positrons). The solid line stands for the thermal Dirac propagator of electrons and positrons and the dashed line for the propagator of the longitudinal ($A_0$) photons. The remaining lines are as in fig. \ref{pNRQEDtreelevel}.}
\label{pNRQEDoneloop}
\end{figure}

In this case the scale $T$ can be integrated out before calculating the spectrum and decay widths, we call the resulting effective theory $pNRQED_{>T}$ (for further explanations about the notation see Appendix \ref{notation}).
In the photon and electron-positron sector this gives rise to the HTL action \cite{Braaten:1991gm}. In the atom sector, the pNRQED Lagrangian gets additional temperature-dependent potentials. At leading order (LO) in $\alpha$ , they arise from the diagram in fig. \ref{pNRQEDtreelevel} upon expanding $E-H$ in the integrals, and have been calculated in \cite{Escobedo:2008sy}
\begin{equation}
\delta V^{(LO)}_T= \frac{\alpha\pi T^2}{3m_\mu}-\frac{4\alpha^2}{3m_\mu^2}\delta^3({\bf r})\left(\frac{1}{\epsilon}+\log\left(\frac{\mu}{2\pi T}\right)+\frac{5}{6}+\log(2\pi)\right)+\mathcal{O}\left(\frac{\alpha r^2E^4}{T}\right) ,
\label{vlo}
\end{equation}
$H\sim E\sim \alpha /r$. Possible ${\cal O} (\alpha r^4 E T^4)$ terms arising from higher orders in the multipole expansion cancel out. Note that the dominant contribution above is a constant mass shift, and the $r$-dependent part is $(m_\mu \alpha^2/n^2T)^2$ suppressed. Hence, vacuum polarization corrections to the photon propagator may compete with the $r$-dependent part of the LO potential displayed above and must be calculated.  
From the diagrams in fig. \ref{pNRQEDoneloop} we obtain the next-to leading order (NLO) in $\alpha$,  
\begin{eqnarray}
\label{vnlo}
&\delta V^{(NLO)}_T= -\frac{3\alpha}{2\pi}\zeta(3)Tm_D^2r^2+\frac{i\alpha Tm_D^2}{6}r^2\left(\frac{1}{\epsilon}+\gamma+\log\pi-\log\frac{T^2}{\mu^2}+\frac{2}{3}-4\log 2-2\frac{\zeta'(2)}{\zeta(2)}\right)+ \\ 
&-\frac{i\alpha m_D^2}{32m_\mu}\left(\frac{1}{\epsilon}+c+2\gamma+2\log\left(\frac{\mu}{T}\right)\right)+\mathcal{O}\left(\frac{\alpha r^2m_D^2E^2}{T},\frac{\alpha r^2 m_D^4}{T}\right), \nonumber
\end{eqnarray}
with $m_D^2=(eT)^2/3$ and $c$ a numerical constant. The computation have been done in dimensional regularization with $\epsilon=(4-d)/2  \to 0$. The first line of this result also appears in an analogous calculation that has already been carried out in the static limit of the QCD case \cite{Brambilla:2008cx}. The second line is subleading. We have displayed it to match the precision of (\ref{vlo}) when $eT\sim E$. In order to eventually check the correct cancellation of the $1/\epsilon$ poles we have calculated analytically the leading IR behavior in
 the appendix \ref{IIA}. The constant $c$ remains unknown.
We see that indeed $\delta V^{(NLO)}_T$ competes in size with $\delta V^{(LO)}_T$, except for the global energy shift given by the first term in (\ref{vlo}). In fact, it provides the dominant term in the potential for the energy of the photon transitions between two states belonging to this case.
The IR divergences arising above are canceled by UV divergences divergences arising from contributions at
smaller scales ($E$, $eT$, $\dots$).
These contributions 
are hard to calculate in the general case because 
 HTL propagators must be used for the ultrasoft photons and the Coulomb potential must be kept unexpanded in the atom propagator. 
At these scales, however, the Bose distribution can be expanded, which simplifies somewhat the calculations, and produces the so called Bose enhancement, see (\ref{relS}) and (\ref{BEe}) below.
The dominant contribution arises from the diagrams of fig. \ref{pNRQCDHTL}.
We have only been able to work out an analytic expressions for the cases $eT \ll E$ and $eT\gg E$, which will be discussed below, and for its UV behavior. Technical details for the latter are shown in the Appendix \ref{IIA}.

Before discussing the two cases which allow to proceed further analytically, we display the energy shift and decay width induced by the temperature-dependent potential (\ref{vlo}) and (\ref{vnlo}) at leading order in perturbation theory,
\bea
&\delta E^T_n=\frac{\alpha\pi T^2}{3m_\mu}-\frac{2\alpha}{3\pi m_\mu^2 }|\phi_n({\bf 0})|^2(\frac{1}{\epsilon}+\log(\frac{\mu}{2\pi T})+\frac{5}{6}+\log(2\pi))\label{ET}\\
&-\frac{3\alpha}{2\pi}\zeta(3)Tm_D^2\langle r^2\rangle_n+\mathcal{O}\left(\frac{\alpha r^2E^4}{T},\frac{\alpha r^2m_D^4}{T}\right) , \nonumber 
\eea
\begin{eqnarray}
&\delta \Gamma^T_n=-\frac{\alpha Tm_D^2\langle r^2\rangle_n }{3}(\frac{1}{\epsilon}+\gamma+\log\pi-\log\frac{T^2}{\mu^2}+\frac{2}{3}-4\log 2-2\frac{\zeta'(2)}{\zeta(2)}) + \label{GT}\\
&+\frac{\alpha m_D^2}{16m_\mu}\left(\frac{1}{\epsilon}+c+2\gamma+2\log\left(\frac{\mu}{T}\right)\right)+\mathcal{O}\left(\frac{\alpha r^2E^4}{T},\frac{\alpha r^2m_D^4}{T}\right), \nonumber
\end{eqnarray}
in which $\langle r^2\rangle_n=\frac{n^2}{2m^2_\mu \alpha^2}[5n^2+1-3l(l+1)]$, $n$ and $l$ being here the principal and angular momentum quantum numbers. The labels $n,m,\dots $ are also used through the paper as a short hand notation for the whole ensemble of quantum numbers of a given Coulomb state, either bound or in the continuum, and $\phi_n({\bf 0})$ is the wave function at the origin.
The contributions above are to be added to the ones coming from lower scales, which we discuss below in two particular cases that share the feature that the lower energy scales are hierarchically ordered, and hence the method of the integration-by-regions can be used \cite{Beneke:1997zp,Smirnov:2002pj}

\begin{itemize}

\item $E \gg eT$

In this case the loop integral is dominated by energy and momentum $\sim E$ for which $eT$ can be treated as a perturbation. At leading order in the HTL expansion we obtain
\be
\label{Egt2}
\delta E^{E}_n=\frac{2\alpha}{3\pi}\sum_m|\langle n|{\bf v}|m\rangle|^2(E_n-E_m)\left(\frac{1}{\epsilon}+\log\left(\frac{\mu}{|E_n-E_m|}\right)+\frac{5}{6}-\gamma+\log(2\pi)\right)-\frac{\alpha\pi Tm_D^2}{3}\langle r^2\rangle_n+\mathcal{O}\left(\frac{\alpha r^2Tm_D^4}{E^2}\right) ,
\ee
\be
\label{DecaygT2}
\delta \Gamma^{E}_n=\frac{4\alpha^3 T}{3n^2}+\frac{\alpha T m_D^2}{3}\sum_m|\langle n|{\bf r}|m\rangle|^2\left(\frac{1}{\epsilon}-2\log\frac{|E_n-E_m|}{\mu}+\frac{11}{3}-\log 4-\gamma+\log(\pi)\right)+\mathcal{O}\left(\frac{\alpha r^2Tm_D^4}{E^2}\right).
\ee
We observe that the leading order infrared divergences appearing at the scale $T$ in (\ref{ET}) and (\ref{GT}) are canceled by the ultraviolet divergences in (\ref{Egt2}) and (\ref{DecaygT2}) respectively. Note that the subleading infrared divergence at the scale $T$, in the second line of (\ref{GT}), is very much suppressed in this case ($m_\mu\alpha^5 \gg \alpha m_D^2/m_\mu$).
A similar calculation for the heavy quarkonium case has been presented in \cite{MilaT2}. Technical details can be found in the appendix (\ref{TEmD}). We only mention here that a collinear region exists that contributes at this order. 

Upon summing up the contributions from both energy regions, namely (\ref{ET}) and (\ref{Egt2}) for the energy and (\ref{GT}) and (\ref{DecaygT2}) for the decay width, finite results are obtained at the desired order,
\begin{eqnarray}
&\delta E_n=\frac{\alpha\pi T^2}{3m_\mu}+\frac{2\alpha}{3\pi}|\phi_n(0)|^2\left(\log\left(\frac{2\pi T}{|E_n|}\right)-\gamma\right)+\frac{2\alpha}{2\pi}\sum_m|\langle n|{\bf v}|m\rangle|^2(E_n-E_m)\log\frac{|E_n|}{|E_n-E_m|} \\
&-\frac{\alpha Tm_D^2\langle r^2\rangle_n}{\pi}\left(\frac{3\zeta(3)}{2}+\frac{\pi^2}{3}\right)+\mathcal{O}\left(\frac{\alpha r^2E^4}{T},\frac{\alpha r^2Tm_D^4}{E^2}\right), \nonumber
\end{eqnarray}
\begin{eqnarray}
&\delta\Gamma_n=\frac{4\alpha^3 T}{3n^2}+\frac{2\alpha Tm_D^2\langle r^2\rangle_n}{3}\left(\log\frac{T}{|E_n|}-\gamma+\frac{3}{2}+\log 2+\frac{\zeta'(2)}{\zeta(2)}\right)-\frac{2\alpha Tm_D^2}{3}\sum_m|\langle n|{\bf r}|m\rangle|^2\log\frac{|E_n-E_m|}{|E_n|} \\
&+\mathcal{O}\left(\frac{\alpha r^2E^4}{T},\frac{\alpha r^2Tm_D^4}{E^2}\right) \nonumber
\end{eqnarray}

\item  $eT \gg E$, 
In this case the loop integral is dominated by energy and momentum $\sim eT$ for which $E$ can be treated as a perturbation. At leading order in the energy expansion we obtain a contribution which is equivalent to adding a new term to the potential that goes like $m_D^2r^2$. 
\be
\label{EgT}
{\delta E^{eT}_n}^{LO}=\frac{\alpha m_D^3}{6}\langle r^2\rangle _n+\mathcal{O}(\alpha^2 r^2m_D^2T) ,
\ee
\be
\label{DecaygT}
{\delta \Gamma^{eT}_n}^{LO}=\frac{\alpha T m_D^2}{3}\langle r^2\rangle_n(\frac{1}{\epsilon}-\gamma+\log \pi +\log\frac{\mu^2}{m_D^2}+\frac{5}{3})+\mathcal{O}(\alpha^2 r^2m_D^2T) .
\ee
We observe that the leading order infrared divergence appearing at the scale $T$ is cancelled by the ultraviolet divergence above. An analogous contribution has also been calculated in the static limit of QCD \cite{Brambilla:2008cx}.

At next-to-leading order in the energy expansion we have restricted ourselves to compute the ultraviolet divergence analytically,
\begin{equation}
\label{DecaygTNLO}
{\delta \Gamma^{eT}_n}^{NLO}=-\frac{\alpha m_D^2}{16m_\mu}\left(\frac{1}{\epsilon}+c^*-2\log\left(\frac{m_D}{\mu}\right)\right)+\mathcal{O}\left(\alpha r^2 E^2T\right)
\end{equation}
It cancels the infrared divergence in the second line of (\ref{vnlo}), as it should ($c^\ast$ is an unknown constant that can be of order $1/\alpha^{1/2}$ because of Bose-enhancement).

\begin{figure}
\includegraphics{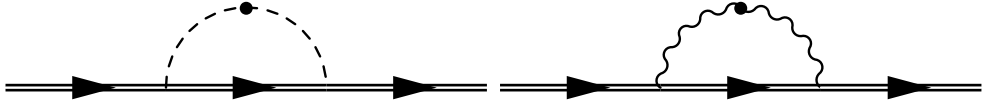}
\caption{Further contributions to the muonic hydrogen atom self-energy when $p\gg T\gg E$. The wavy line and the dashed line with a blob are the HTL propagators for the transverse and longitudinal photons respectively.}
\label{pNRQCDHTL}
\end{figure}

Fortunately the contribution of the loop integral for energy and momenta $\sim E$ is subleading. The calculation at that scale may 
even require non-perturbative techniques 
if $E$  gets close to the scale $e^2 T$ \cite{Lebellac,Bodeker:1998de,Laine:2007qy}. 

Summing up the contributions from the $T$ energy region and from the $m_D$ energy region, namely (\ref{ET}) and (\ref{EgT}) for the energy and (\ref{GT}) and (\ref{DecaygT}) for the decay width, the leading thermal effects for this situation are obtained,
\begin{equation}
\delta E_n=\frac{\alpha\pi T^2}{3m_\mu}+\frac{\alpha m_D^3\langle r^2\rangle_n}{6}+\mathcal{O}\left(
\alpha r^2E^2T\right)
\end{equation}
\begin{equation}
\delta\Gamma_n=\frac{2\alpha Tm_D^2\langle r^2\rangle_n}{3}\left(\log\frac{T}{m_D}-\gamma+\frac{1}{2}+2\log 2+\frac{\zeta'(2)}{\zeta(2)}\right)-\frac{2\alpha m_D^2}{16m_\mu}\left(\log\frac{T}{m_D}+\frac{c^*-c}{2}-\gamma\right)+\mathcal{O}\left(
\alpha r^2E^2T\right)
\end{equation}
\end{itemize}

\subsection{$T\sim p$}

Since $m_\mu\gg T$ still holds, our starting point is the NRQED Lagrangian at $T=0$ \cite{Caswell:1985ui} (this is correct up to exponentially small contributions $\sim e^{-m_\mu / T}$). 

\begin{eqnarray}
\mathcal{L}=\psi^+(iD^0+\frac{{\bf D}^2}{2m_\mu}+\frac{{\bf D}^4}{8m_\mu^3}+c_F e\frac{{ \bfsigma}{\bf B}}{2m_\mu}+c_D e\frac{|{ \bfnabla}{\bf E}|}{8m_\mu^2}+\\
+ic_S e\frac{{\bfsigma}({\bf D}\times{\bf E}-{\bf E}\times{\bf D})}{8m_\mu^2})\psi+N^+iD^0N-\frac{1}{4}F_{\mu\nu}F^{\mu\nu}+\frac{d_2}{m_\mu^2}F_{\mu\nu}D^2F^{\mu\nu}+\bar{e}i\gamma^\mu D_\mu e . \nonumber
\end{eqnarray}
where $\psi$ and $N$ are the muon and proton Pauli-spinor fields respectively.

\begin{figure}
\includegraphics{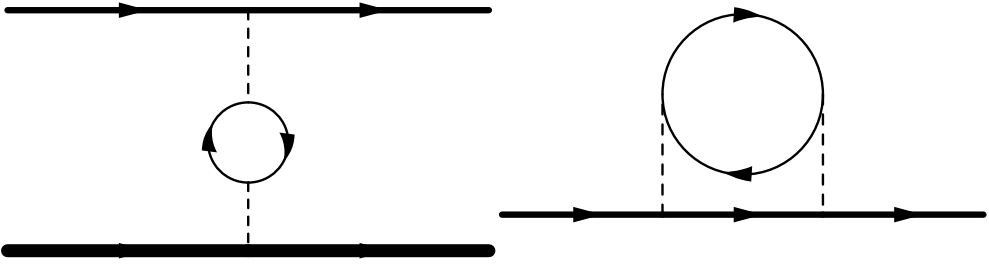}
\caption{Leading correction to the Coulomb potential due to vacuum polarization when $T\lesssim p$. The thick and the extra-thick solid lines stand for the non-relativistic propagator of the muon and the static propagator of the proton respectively. The remaining lines are as in fig. \ref{pNRQEDoneloop}. There is also a leading order correction coming from an extra diagram, which is obtained from the second one by changing the muon line by a proton line.}
\label{NRQED}
\end{figure}

Since $T\sim p\gg E$, we can integrate out the scales $T$ and $p$, which leads to what we call $pNRQED_T$, the suitable effective theory for the scale $E$, which is similar to the one that has already been introduced in section II.A, with the only difference that now $T$ is of the same order as the cut off of the effective theory. In the photon and electron-positron sectors we have the standard HTL. In the atom sector temperature-dependent potentials are induced. Recall that at the scale $T$ there is no enhancement and vacuum polarization effects due to electron-positron pairs are always suppressed by $\alpha$. Hence, the leading potential will still be the Coulomb potential but the first $\alpha$ correction to it will already be temperature-dependent. This is given by the diagram in fig. \ref{NRQED}.

The temperature-dependent part of the potential induced by the diagram in fig. \ref{NRQED} is both UV and IR safe in momentum space. However, when it is Fourier transformed to coordinate space an IR divergence is encountered. The calculation is rather involved so we only display here the final result, which can be given in terms of one-parameter integrals of special functions. Details are given in the Appendix \ref{IIIA} (the formulas in that appendix have to be used setting $m_e=0$ for this case). We obtain,

\bea
\label{potm0}
&\delta V_r=-\frac{\alpha m_D^2r}{4}-\frac{3\alpha}{2\pi}\zeta(3)Tm_D^2r^2+ \\
&+\frac{\alpha m_D^2}{4\pi^2T^2r}\int_0^\infty\frac{\,du}{u(e^u+1)}(-4-4\rho^2u^2+(\rho^2u^2+4)\cos(\rho u)+\nonumber \\
&+\rho u\sin(\rho u)+(6\rho u+\rho^3u^3)Si(\rho u))+\nonumber \\
&+\frac{i\alpha m_D^2Tr^2}{6}(\frac{1}{\epsilon}+\gamma+\log\pi+2\log(r\mu)-1)+\nonumber \\
&-\frac{i3\alpha m_D^2}{2\pi^2 T}(\frac{1}{2}-\log(rT)-\log\pi)+\nonumber \\
&+\frac{i3\alpha m_D^2}{\pi^2T^2r}\int_0^\infty\frac{\,du}{u^4}\sin(\rho u)(Li_2(-e^{u})+u\log(1+e^u)+\frac{\pi^2}{12}-\frac{u^2}{4})+\mathcal{O}(\alpha^3 T) , \nonumber
\eea
where $\rho=2rT$ and $Si$ stands for Sine Integral
\be
Si(z)=\int_0^z\frac{\sin t}{t}\,dt .
\ee

The LO energy correction is obtained by computing the expectation value of the potential of (\ref{potm0}) for the desired state and adding the ultrasoft contribution. The calculation in $pNRQED_T$ is identical to one carried out in the second case of sect. A (since $E\sim m_\mu \alpha^2$, $T\sim p\sim m_\mu \alpha$, we have $eT \gg E$). Hence the outcome can be directly read off
equations (\ref{EgT}) and (\ref{DecaygT}).
 Notice that the infrared divergence in (\ref{potm0}) at first order in quantum mechanic perturbation theory, induces a contribution that cancels out the ultraviolet divergence in (\ref{DecaygT}), as it should. Also, the contribution from integrating out the scale $m_D$ can be encoded in a correction to the potential, summing it up to $\delta V_r$ the following finite result is obtained,
 
\bea
&\delta V=-\frac{\alpha m_D^2r}{4}-\frac{3\alpha}{2\pi}\zeta(3)Tm_D^2r^2+\frac{\alpha m_D^3r^2}{6}+ \\
&+\frac{\alpha m_D^2}{4\pi^2T^2r}\int_0^\infty\frac{\,du}{u(e^u+1)}(-4-4\rho^2u^2+(\rho^2u^2+4)\cos(\rho u)+\nonumber \\
&+\rho u\sin(\rho u)+(6\rho u+\rho^3u^3)Si(\rho u))+\nonumber \\
&-\frac{i\alpha m_D^2Tr^2}{3}(-\log(rm_D)+\frac{4}{3}-\gamma)+\nonumber \\
&-\frac{i3\alpha m_D^2}{2\pi^2 T}(\frac{1}{2}-\log(rT)-\log\pi)+\nonumber \\
&+\frac{i3\alpha m_D^2}{\pi^2T^2r}\int_0^\infty\frac{\,du}{u^4}\sin(\rho u)(Li_2(-e^{u})+u\log(1+e^u)+\frac{\pi^2}{12}-\frac{u^2}{4})+\mathcal{O}(\alpha^3 T) , \nonumber
\eea
and hence, at first order in perturbation theory,
\begin{eqnarray}
&\delta E_n=\langle n|\Re\delta V|n\rangle \label{EG}\\
&\delta\Gamma_n=-2\langle n|\Im\delta V|n\rangle\nonumber
\end{eqnarray}

\subsection{$T\gg p$}

Since $m_\mu\gg T$ still holds, we can also start from NRQED at $T=0$. Now we may proceed by sequentially integrating out first the scale $T$ and next the scale $p$. After integration of the scale $T$ we get an effective theory which consist of HTL contributions in the photon and electron-positron sector, and of NRQED with temperature-dependent matching coefficients in the atom sector. This $NRQED_{>T}$ in the atom sector is identical to the one that we have in the hydrogen atom case \cite{Escobedo:2008sy}
, up to order $\alpha$ corrections induced by the electron-positron vacuum polarization. 

The next step is to integrate out the energy scale $p$, namely to match $NRQED_{>T}$ to what will be called $pNRQED_{<T}$, which is expected to produce temperature-dependent potentials. These potentials must be calculated using HTL photon propagators. Let us separate the two following cases:

\begin{itemize}
\item $eT \sim p$, 

In this case the computations can be carried out as in section V.B of \cite{Escobedo:2008sy}. The relevant diagram is similar to fig. \ref{NRQED}, but instead of a photon propagator with a self-energy insertion we would have to use the tree level HTL photon propagator (fig. \ref{NRQEDHTL}). The only difference with respect to \cite{Escobedo:2008sy} is due to the fact that the electron-positron pairs that generated the HTL photon propagators
 are now taken to be massless, so, in fact, the outcome is simpler: the non-trivial function $g(m_e\beta )$ reduces to
$\frac{\pi m_D^2}{16\alpha T^2}$ .
We obtain then the following leading order potential
\be
V(r,T)=-\frac{\alpha e^{-m_Dr}}{r}-\alpha m_D+i\alpha T\phi(m_Dr)+\mathcal{O}\left(\frac{\alpha T^2}{m_\mu}\right) ,
\ee
where
\begin{equation}
\phi(x)=2\int_0^\infty\frac{\,dzz}{(z^2+1)^2}\left[\frac{\sin(zx)}{zx}-1\right].
\end{equation}
This potential coincides with the one first obtained in \cite{Laine:2006ns} for QCD (up to trivial changes in color factors made explicit in \cite{Escobedo:2008sy}).
As in the hydrogen atom case, we can use this result in order to estimate the dissociation temperature,
which is $T_d\sim m_\mu\alpha^{2/3}/\ln^{1/3}\alpha$, as anticipated in \cite{Escobedo:2008sy} for the QCD case.

\begin{figure}
\includegraphics{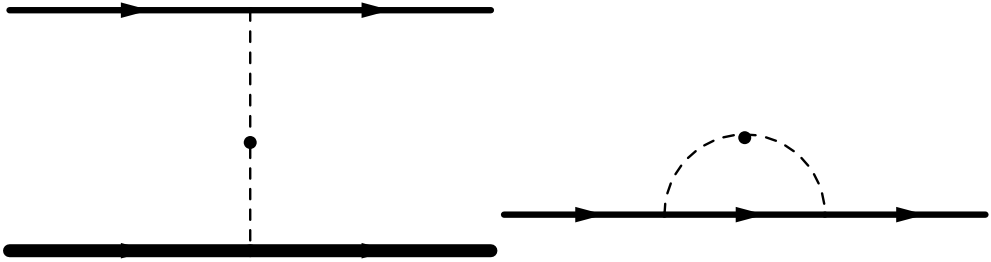}
\caption{Leading correction to the Coulomb potential due to vacuum polarization when $T\gg p$. 
The lines are as in figs. \ref{NRQED} and \ref{pNRQCDHTL}.}
\label{NRQEDHTL}
\end{figure}

\item $eT \gg p$, 

In this case $T\gg T_d$ always holds. Hence, the imaginary part of the potential is bigger than the real one, so 
it does not make much sense to speak about
 bound states anymore.

\end{itemize}

\section{$m_e\not=0$ case}

We address now the actual case of a non-vanishing electron mass. Although the real muonic hydrogen is 
not as close to heavy quarkonium systems as the ideal one ($m_e=0$), it may still be useful to learn about certain aspects of it. In particular about the r\^ ole of a finite charm mass in the bottomonium system, which is analogous to that of a finite electron mass in muonic hydrogen \cite{Eiras:2000rh}. This case may then shed light on the effects of the charm quark mass in bottomonium at finite temperature, specially when charm quarks are thermalized.
Irrespective of that, muonic hydrogen is a real system that appears in nature, which nowadays is produced in large samples \cite{PSI}, and, therefore, our results may eventually be checked against experiment.

For actual muonic hydrogen $m_e \sim p$ for the lower lying states ($n=1,2$), whereas for the remaining states ($n \ge 3$) one may safely consider $m_e \gg p$ \cite{Eiras:2000rh}. Let us then analyse these two cases separately.  

\subsection{Lower lying states ($n=1,2$)}

As mention above, these two states fulfill $p\sim m_e$, and hence relativistic electron-positron pairs must be integrated at the same time as the momentum transfer $p$ is. Let us see in the following sections how this is carried out depending on what the temperature is.

\subsubsection{$p\gg T$}

Like in the massless case our starting point can be pNRQED. However, now, due to the fact that $m_e \sim p$ (rather than $m_e=0$),
the electron-positron pairs have already been integrated out when calculating the potentials, and hence are not active anymore. The situation is then totally analogous to the hydrogen atom, the only difference being that the potentials get ${\cal O}(\alpha)$ corrections due to virtual electron-positron pairs, the most important of which is the Uehling potential. In other words, the thermal bath contains neither electrons nor positrons, so the thermal effects are only due to the photons which do not distinguish between electrons and muons. Hence the results concerning this case can be read off section III of ref. \cite{Escobedo:2008sy} by making $m\to m_\mu$ [up to ${\cal O}(\alpha)$ corrections].

\subsubsection{$T\sim p$}

Again, like in the massless case, we can start with NRQED.
 The scales $T$ and $m_e$ must integrated out at the same time as the energy scale $p$. In the photon sector, which is not sensitive to the scale $p$, we get the mass dependent HTL action (see section V.A.1 of \cite{Escobedo:2008sy}). In the electron and positron sectors, which are not sensitive to the scale $p$ either, we get a NRQED$_T$ Lagrangian for each of these particles (see section V.A.2 of \cite{Escobedo:2008sy}). In the atom sector, 
the potentials depend now on both temperature and the electron mass, except for the leading Coulomb potential. The most important correction is a kind of temperature-dependent Uehling potential, which is obtained from the diagram in fig. \ref{NRQED},
\bea
\label{potmno0}
&\delta V_r=-\frac{4\alpha^2f(m_e\beta)m_e^2r}{\pi}-\frac{2\alpha^2}{\pi r}\int_0^\infty\frac{\,du}{\sqrt{u^2+1}(e^{\beta m_e\sqrt{u^2+1}}+1)}(1-\cos(\sigma u)-\sigma uSi(\sigma u))+\\
&+\frac{\alpha^2}{3\pi r}\int_0^\infty\frac{\,du\sqrt{u^2+1}}{u^2(e^{\beta m_e\sqrt{u^2+1}}+1)}\left(2-3\sigma^2u^2+(\sigma^2u^2-2)\cos(\sigma u)+\right.\nonumber \\
&\left.+\sigma u\sin(\sigma u)+\sigma^3u^3Si(\sigma u)\right)-\frac{\alpha}{\pi}Tm_D^2r^2(\beta m_e)^3\int_0^\infty\frac{\,du u\sqrt{u^2+1}}{e^{\beta m_e\sqrt{u^2+1}}+1}+\nonumber \\
&+\frac{i8\alpha^2T^3g(m_e\beta)r^2}{3\pi}(\frac{1}{\epsilon}+\gamma+\log\pi+\log(r\mu)^2-1)-\nonumber \\
&-\frac{i4\alpha^2 T}{\pi(e^{\beta m_e}+1)}\left(\frac{1}{2}-\log(rT)-\log 2-(e^{\beta m_e}+1)\int_0^\infty\frac{\,du}{u(e^{\beta m_e\sqrt{u^2+1}}+1)}+\int_0^\infty\frac{\,due^{-\beta m_eu}}{u}\right)+\nonumber \\
&+\frac{i4\alpha^2T^3}{\pi rm_e^3}\int_0^\infty\frac{\,du}{u^4}\sin(\sigma u)\left(Li_{2}(-e^{\beta m_e\sqrt{u^2+1}})+(\beta m_e)\sqrt{u^2+1}\log(1+e^{\beta m_e\sqrt{u^2+1}})\right.+\nonumber \\
&\left.+\frac{\pi^2}{6}-\frac{(\beta m_e)^2}{2}(u^2+1)-g(m_e\beta)+\frac{(\beta m_e)^2u^2}{2(e^{\beta m_e}+1)}\right)+\nonumber \\
&+\frac{i4\alpha^2T}{\pi}\int_0^\infty\frac{\,du}{u^3}\left(\frac{Sinc(\sigma u)-1}{e^{\beta m_e\sqrt{u^2+1}}+1}-\frac{Sinc(\sigma u)-e^{-8\beta^3m_e^3u^3}}{e^{\beta m_e}+1}\right)+\frac{i4\alpha^2 m_e^2Tr^2}{3\pi(e^{\beta m_e}+1)}(\frac{1}{\epsilon}-1+\gamma+2\log(r\mu)+\log\pi)- \nonumber \\
&-\frac{i16\alpha^2m_e^2}{3\pi T(e^{\beta m_e}+1)}\Gamma(-2/3)+\mathcal{O}(\alpha^3 T) , \nonumber
\eea
where $\beta=1/T$, $\sigma=2m_er$ and 
\bea
f(m_e\beta )&=& \int_0^\infty\,dx\frac{x^2}{\sqrt{x^2+1}(e^{\beta m_e\sqrt{x^2+1}}+1)}, \\
g(m_e\beta )&=& \beta^2m_e^2\int_0^\infty\,dx\frac{x}{e^{\beta m_e\sqrt{x^2+1}}+1}.
\eea
Further expressions for these functions can be found in the Appendix B of \cite{Escobedo:2008sy}. The computations that lead to (\ref{potmno0}) are carried out in Appendix \ref{IIIA}. 

Notice that  (\ref{potmno0}) has two infrared divergences, which, as in the $m_e=0$ case, arise when the Fourier transform of the momentum space potential is taken, in order to get the coordinate space potential. The IR divergence in the fourth line of (\ref{potmno0}) is similar to the one that appears in equation (\ref{potm0}) for the massless case (with $T^2g(m_e\beta)$ instead of $m_D^2$). The IR divergence in the second last line of (\ref{potmno0}), however, is proportional to $\frac{m_e^2}{e^{\beta m_e}+1}$ 
, and hence distinct of the $m_e=0$ case. It emerges from a region in which not only the three-momentum transfer is small but also the component of the three-momentum of the electron-positron pair in the loop parallel to the momentum transfer is small. In either case, these IR divergences should cancel against UV divergences in the pNRQED calculation.

The relevant diagram in the pNRQED calculation is again fig. \ref{pNRQCDHTL}, in which the photon line must be understood as the mass-dependent HTL propagator (see Appendix \ref{HTLme}). In the dominant contribution to this diagram, $E-H$ in the atom propagator can be treated as a perturbation (recall that $E-H \sim m_\mu \alpha^2 \ll eT$). Then using the fact that $\Delta_{11}(k^0, {\bf k})$ is symmetric with respect to $k^0\to -k^0$, we obtain

\be
\label{meeT}
\delta E^{eT}_n=\frac{\alpha m_D^3}{6}\langle r^2\rangle_n+\mathcal{O}(\alpha r^2Em_D^2) ,
\ee
\bea
\label{DecaymgT}
&\delta \Gamma_n^{eT}=\left[\frac{16\alpha^2T^3}{3\pi}g(m_e\beta)(\frac{1}{\epsilon}-\gamma+\log\pi+\log\frac{\mu^2}{m_D^2}+\frac{5}{3})+ \right.\\
& \left. +\frac{8\alpha^2m_e^2T}{3\pi(e^{\beta m_e}+1)}(\frac{1}{\epsilon}-2\log\frac{m_D}{\mu}+\frac{5}{3}+\log(4\pi)-\gamma-2\log 2)\right]\langle r^2\rangle_n+\mathcal{O}(\alpha r^2Em_D^2). \nonumber
\eea
We see that indeed the UV divergences above cancel those of (\ref{potmno0}), as expected.
There is a subtle point in this calculation, however, which we discuss in the Appendix \ref{HTLme}, that must be correctly dealt with in order to get the UV divergence of the last line (that cancels the IR divergence in the second last line of (\ref{potmno0})). As in the massless case, the contribution from the scale $m_D$ can be encoded in a correction to the potential, and this summed to $\delta V_r$,
\bea
&\delta V=-\frac{4\alpha^2f(m_e\beta)m_e^2r}{\pi}-\frac{2\alpha^2}{\pi r}\int_0^\infty\frac{\,du}{\sqrt{u^2+1}(e^{\beta m_e\sqrt{u^2+1}}+1)}(1-\cos(\sigma u)-\sigma uSi(\sigma u))+\label{potfin}\\
&+\frac{\alpha^2}{3\pi r}\int_0^\infty\frac{\,du\sqrt{u^2+1}}{u^2(e^{\beta m_e\sqrt{u^2+1}}+1)}\left(2-3\sigma^2u^2+(\sigma^2u^2-2)\cos(\sigma u)+\right.\nonumber \\
&\left.+\sigma u\sin(\sigma u)+\sigma^3u^3Si(\sigma u)\right)-\frac{\alpha}{\pi}Tm_D^2r^2(\beta m_e)^3\int_0^\infty\frac{\,du u\sqrt{u^2+1}}{e^{\beta m_e\sqrt{u^2+1}}+1}+\nonumber \\
&+\frac{\alpha m_D^3r^2}{6}-\frac{i16\alpha^2T^3g(m_e\beta)r^2}{3\pi}(-\log(rm_D)+\frac{4}{3}-\gamma)-\nonumber \\
&-\frac{i4\alpha^2 T}{\pi(e^{\beta m_e}+1)}\left(\frac{1}{2}-\log(rT)-\log 2-(e^{\beta m_e}+1)\int_0^\infty\frac{\,du}{u(e^{\beta m_e\sqrt{u^2+1}}+1)}+\int_0^\infty\frac{\,due^{-\beta m_eu}}{u}\right)+\nonumber \\
&+\frac{i4\alpha^2}{\pi r}\int_0^\infty\frac{\,du}{u^4}\sin(\sigma u)\left(Li_{2}(-e^{\beta m_e\sqrt{u^2+1}})+\beta m_e\sqrt{u^2+1}\log(1+e^{\beta m_e\sqrt{u^2+1}})+\right.\nonumber \\
&\left.+\frac{\pi^2}{6}-\frac{(\beta m_e)^2}{2}(u^2+1)-g(m_e\beta)+\frac{(\beta m_e)^2u^2}{2(e^{\beta m_e}+1)}\right)+\nonumber \\
&+\frac{i4\alpha^2T}{\pi}\int_0^\infty\frac{\,du}{u^3}\left(\frac{Sinc(\sigma u)-1}{e^{\beta m_e\sqrt{u^2+1}}+1}-\frac{Sinc(\sigma u)-e^{-8\beta^3m_e^3u^3}}{e^{\beta m_e}+1}\right)-\frac{i8\alpha^2 m_e^2Tr^2}{3\pi(e^{\beta m_e}+1)}(-\log(rm_D)+\frac{4}{3}-\gamma)- \nonumber \\
&-\frac{i16\alpha^2m_e^2}{3\pi T(e^{\beta m_e}+1)}\Gamma(-2/3)+\mathcal{O}(\alpha^3 T) . \nonumber
\eea
The energy shift and the decay width at first order in perturbation theory can be obtained from (\ref{EG}).
\subsubsection{$T\gg p$}
\label{llTggp}

This case is very similar to the $m_e=0$ one. We start with NRQED and integrate out the scale $T$ first. Since $T\gg m_e$, the mass-dependent HTL propagators may be expanded in 
$m_e/T$, and hence become the usual HTL propagators with next-to-leading order contributions due to the non-vanishing electron mass . 
Hence the finite mass effects do not affect the gross features of the system. In particular, in the $eT \sim p$ case the dissociation temperature will be  similar to the one in the massless case. For $eT \gg p$, like in the massless case, no bound state is expected to survive.

\subsection{Higher energy states ($n\ge 3$)}

As mentioned before, these states fulfill
 $m_e \gg p \gg E$, and hence relativistic electron-positron pairs may be integrated out before the momentum transfer $p$ is. Let us see in the following sections how this is carried out depending on what the temperature is.

\subsubsection{$m_e\gg T$}

In this case we can start with a NRQED Lagrangian for the muon in which the relativistic electron-positron pairs have already been integrated out, which gives rise to $1/m_e^2$ corrections to the Maxwell Lagrangian. Since there are neither electrons nor positrons in the thermal bath, the different situations coincide with those of the hydrogen atom, and the relevant expressions can be read off sections III and IV of \cite{Escobedo:2008sy} by making $m\to m_\mu$ (up to $1/m_e^2$ corrections).

\subsubsection{$m_e \lesssim T$}
\label{hemelesssimT}

In this case we can start with a NRQED Lagrangian for the muon keeping the relativistic Dirac Lagrangian for the electron in it.
 When $m_e\sim T$ both scales must be integrated out at the same time. In the photon sector, a mass-dependent HTL Lagrangian is induced (see section V.A.1 of \cite{Escobedo:2008sy}). In the electron and positron sectors a temperature-dependent NRQED$_T$ Lagrangian for each particle is induced (see section V.A.1 of \cite{Escobedo:2008sy}). In the muon sector, a temperature-dependent NRQED$_{>T}$ Lagrangian is also induced. At lower orders, it can be obtained from the diagrams of section IV of \cite{Escobedo:2008sy}, together with diagrams containing an electron-positron loop. The most important effect is the appearance of two mass shifts, one $\sim \alpha T^2/m_\mu$ from the diagram (38) of \cite{Escobedo:2008sy} and the other one $\sim \alpha^2 m_e$ from the second diagram in fig. \ref{NRQED}. In the proton sector an analogous mass shift $\sim \alpha^2 m_e$ occurs, that is due to a diagram obtained from the previous one by changing the muon line by a proton line.

The next step is to integrate out the scale $p$, namely matching to pNRQED using the HTL Lagrangian. This has already been done in section V.B of \cite{Escobedo:2008sy}. At leading order, this produces a temperature-dependent potential and further mass shifts, which can be read of a corrected version of formulas (58) and (60) in that reference (making $m\to m_e$). The origin of the corrections is discussed in Appendix \ref{HTLme}, and boils down to a simple replacement, see (\ref{correction}). Both the potential and the mass shift contain an imaginary part, which becomes more important than the real part starting at some temperature $T_d$, which we call dissociation temperature. In the following section
the dissociation temperatures will be estimate for several states.
 
For $T\gg m_e$ none of the higher energy states exists anymore, since the dissociation temperatures fulfill $m_e > T$, see next section. 

\subsection{Dissociation temperatures, level shifts and decay widths}

The dissociation temperatures will be estimated in a similar way as they were in the hydrogen atom case \cite{Escobedo:2008sy}. 
This is by identifying the momentum scale for which the real and imaginary part of the momentum space potential are equal, 
$p\sim (16\alpha)^{1/3}(g(m_e\beta)+(m_e\beta)^2n_F(m_e\beta)/2)^{1/3} T=:m_d$, and equating it to the typical momentum transfer 
in the muonic hydrogen atom\footnote{Note that in ref. \cite{Escobedo:2008sy} the typical momentum of the electron, $p\sim  m_e \alpha /n$, 
was used rather than the typical momentum transfer, $p\sim  m_e \alpha /n^2$, as we have adopted here. For the lower lying states the order 
of magnitude estimates does not differ much, but for higher energy ones may differ considerably. We have also identified an error, 
that affects the dissociation temperatures, in the computation of the potential (57) of \cite{Escobedo:2008sy}, 
which is explained and corrected in Appendix \ref{HTLme}. 
We reproduce here the table I of ref. \cite{Escobedo:2008sy} in table II with the current choice of $p$ and the correct version of the potential 
for the sake of comparison.}, $p\sim  m_\mu \alpha /n^2$. The results are displayed in table I. 
Table II shows the same results for the hydrogen atom. In order to carry out the estimates we have use the formulas of Appendix \ref{HTLme}, which are meant for higher excited states ($T\sim m_e\gg p$), for the case of the lower lying states, instead of the $m_e=0$ formulas. This is indeed legitimated: We are just not taking advantage of the fact that $T_d\gg m_e$ for these states, which produces the simplifications discussed in section \ref{llTggp}.

\begin{table}[htb]
\makebox[6cm]{\phantom b}
\begin{center}
\begin{tabular}{|c|c|c|c|}
\hline
n & $T_d$ (MeV) & $m_D$ (MeV) & $m_d$ (MeV) \\
\hline
1 & 1.7 & 0.16 & 0.77 \\
2 & 0.41 & 0.036 & 0.19 \\
3 & 0.19 & 0.012 & 0.086 \\
4 & 0.13 & 0.0056 & 0.048 \\
5 & 0.10 & 0.0030 & 0.031 \\
\hline
\end{tabular}
\end{center}
\caption{Dissociation temperature for the lower lying states of muonic hydrogen}\end{table}

\begin{table}[htb]
\makebox[6cm]{\phantom b}
\begin{center}
\begin{tabular}{|c|c|c|c|}
\hline
n & $T_d$ (KeV) & $m_D$ (KeV) & $m_d$ (KeV) \\
\hline
1 & 49 & 0.15 & 3.7 \\
2 & 36 & 0.020 & 0.93 \\
3 & 31 & 0.0061 & 0.41 \\
4 & 28 & 0.0025 & 0.23 \\
5 & 26 & 0.0013 & 0.15 \\
\hline
\end{tabular}
\end{center}
\caption{Dissociation temperature for the lower lying states of hydrogen}
\end{table}

\begin{figure}
\includegraphics{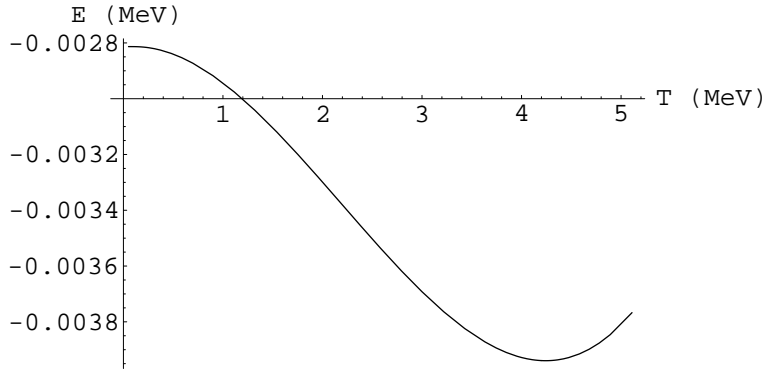}
\caption{$E$ vs $T$ for the 1S state. This result was computed using the assumption that $\frac{1}{r}\gg m_D$, so we expect important deviations from the real energy starting at $T\sim 4 MeV$}
\label{grafica1s}
\end{figure}

\begin{figure}
\includegraphics{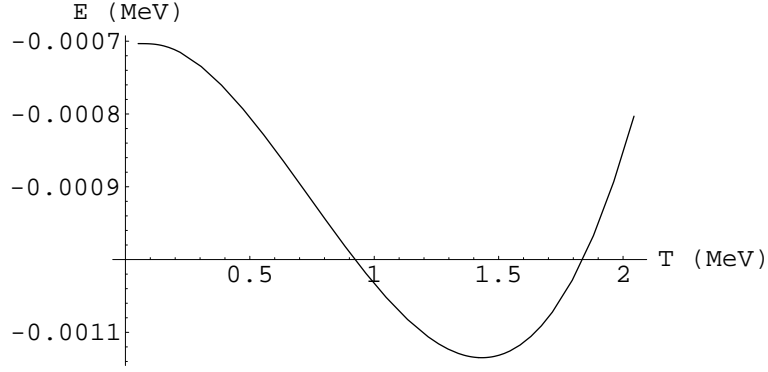}
\caption{$E$ vs $T$ for the 2P state. For the 2P state the typical radius is four times bigger than for the 1S state, so we expect important deviations from the real energy starting at $T\sim 1 MeV$}
\label{grafica2p}
\end{figure}

\begin{figure}
\includegraphics{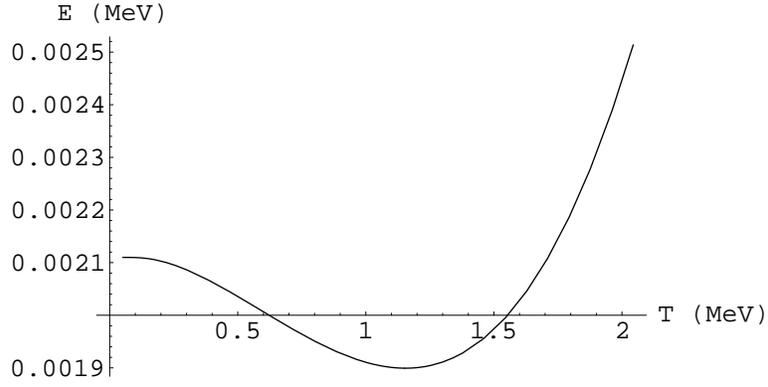}
\caption{$K_\alpha$ transition vs $T$. This result is reliable until $T\sim 1 MeV$. Note that this temperature is twice what we predicted for the dissociation temperature of the 2P state}
\label{lineak}
\end{figure}

From the results in table I, we see that only the lowest lying states $n=1,2$ survive at temperatures of the electron-positron plasma $m_e\lesssim T$  
Hence, only transitions between these levels might be observed in an eventual experiment. We shall focus on the experimentally prominent $K_\alpha$ transition \cite{Klines}. We display our results for the energies of the $1S$ and $2P$ states, and for the energy of the $K_\alpha$ transition as a function of  temperature in the range $T\in (2, 0.05) MeV.$ in fig. \ref{grafica1s}, fig. \ref{grafica2p} and fig. \ref{lineak} respectively. The calculations have been carried out numerically, using first order perturbation theory for the potential (\ref{potfin}).

\begin{figure}
\includegraphics{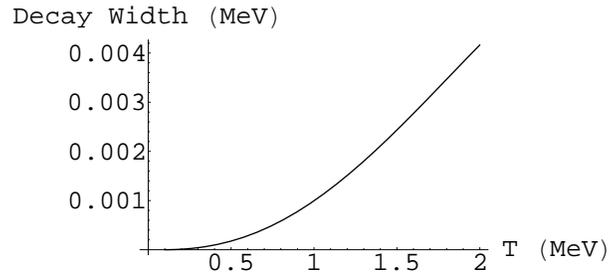}
\caption{Decay width for the 1S state. As for the binding energy this has been computed using the assumption that $\frac{1}{r}\gg m_D$}
\label{de1s}
\end{figure}

Another observable that we can predict with our results is the decay width. In fig. \ref{de1s} we show the decay width for the 1S state as a function of the temperature. Comparing fig. \ref{grafica1s} and fig. \ref{de1s} it can be seen that $T_d\sim 1.7 MeV$ is the temperature that makes the decay width of the same magnitude as the binding energy.

\section{Discussion and Conclusions}
In this paper we have discussed the properties of muonic hydrogen in a thermal bath, which may consist not only of blackbody radiation but also of an electron-positron plasma. We have further developed the effective theory techniques for bound state systems at finite temperature initiated in \cite{Escobedo:2008sy}, in particular the application of dimensional regularization to the factorization of the various scales in the system. They facilitate enormously the organisation of the calculation. For instance, they make apparent when Coulomb or HTL resummations are necessary and when they are not. In addition, both partial and final results are naturally obtained as a series of small scales over large ones, thus providing a good control on the systematics.

We have discussed two cases. We have first addressed the academic case of muonic hydrogen with a vanishing electron mass, which turns out to be  closer to heavy quarkonium states than the actual case with a non-vanishing electron mass, that we have addressed next. All the thermal modifications we have found turn out to be spin independent. 

In the zero electron mass case, we have studied how the effects of vacuum polarization modify the picture that we encounter in normal hydrogen \cite{Escobedo:2008sy}. The modifications turn out to be important when the temperature is larger than the binding energy.
For instance, they would give the leading order contribution to a hypothetical $K_\alpha$ transition for high enough temperatures (\ref{vlo}). For temperatures below dissociation, we 
have presented the leading order, and selected next-to-leading order, thermal corrections to the binding energy and decay width.

In the actual electron mass case, 
muonic hydrogen behaves very much the same as hydrogen for temperatures below the electron mass. For temperatures larger or of the order of the electron mass the vacuum polarization effects are sizable, and, at some point, make the bound states dissociate.
We display in table I the dissociation temperature for the lower laying states. We have also calculated the thermal modifications to a number of observables before dissociation occurred. For instance, we plot the dependence of the $K_\alpha$ transition on temperature in fig. \ref{lineak}, which  could be tested experimentally in the future \cite{PSI}.

We close with a concrete application to the heavy quarkonium case. As we have mentioned before, the way a finite electron mass affects muonic hydrogen 
is similar to the way a finite charm quark mass affects bottomonium \cite{Eiras:2000rh}. Since this should also be the case at finite temperature, we can easily translate to the QCD case the results for the dissociation temperature of muonic hydrogen, which we show in table III.

\begin{table}[htb]
\makebox[6cm]{\phantom b}
\begin{center}
\begin{tabular}{|c|c|}
\hline
$m_c$ (MeV) & $T_d$ (MeV) \\
\hline
$\infty$ & 480 \\
5000 & 480 \\
2500 & 460 \\
1200 & 440 \\
0 & 420 \\
\hline
\end{tabular}
\end{center}
\caption{Dissociation temperature for Upsilon (1S) for different values of the charm mass. The $n_f=3$ light quark masses are set to zero. We use as an input the values of the Bohr radius and $\Lambda_{QCD}$ found in table 2.1 of \cite{Pinedathesis}. The values of these parameters for $n_F=3$ are used for all values of $m_c$ except for $m_c=0$, where we use the ones for $n_F=4$}
\end{table}

\begin{acknowledgments}
We acknowledge financial support from the RTN Flavianet MRTN-CT-2006-035482 (EU), the FPA2007-60275/MEC grant (Spain) and the 2009SGR502 CUR grant (Catalonia). J.S also acknowledges financial support form the ECRI HadronPhysics2 (Grant Agreement n. 227431) (EU), the FPA2007-66665-C02-01/MEC grant,and the Consolider Ingenio program CPAN CSD2007-00042 (Spain). M.A.E. has also been supported by a MEC FPU (Spain).
\end{acknowledgments}

\appendix

\section{Notation for the different effective field theories}
\label{notation}
At zero temperature there are three different energy scales for non-relativistic bound states. These are the hard scale (for muonic hydrogen $m_\mu$), the soft scale $m_\mu\alpha$ and the ultrasoft scale $m_\mu\alpha^2$. Moreover, a finite temperature system also has a different energy scale as $T$, $eT$, $e^2T$...
This makes it hard to find a comprehensible notation for all the effective field theories that may arise from integrating out the different degrees of freedom. In this paper we have used the following notation. Basically we name the effective field theories as one would do for zero temperature, and we encode the temperature information in a subindex. The subindex $T$ means that the temperature has been integrated out, the subindex $m_D$ means that also the scale $eT$ has been integrated out. Since the matching coefficients of the effective field theory that we obtain after integrating out $m_\mu$ ($p$) and $T$ is not the same if $m_\mu \, (p) \sim T$ or if $m_\mu \, (p) >>T$ we include a symbol $<$, $>$ or blank depending of the relation between these scales. For example, if we are in $m_\mu \, (p) \sim T$ we will arrive to  $NRQED_T$ ($pNRQED_T$), but if $m_\mu \, (p)>>T$ we reach $NRQED_{>T}$ ($pNRQED_{>T}$) because $T$ is smaller than the energy cutoff of $NRQED$ ($pNRQED$).

\section{Basic formulas}
\label{review}
In this appendix we display 
a number of formulas of the real-time formalism that are relevant for the paper. Our notation closely follows ref. \cite{Thoma:2000dc}. Recall that  
in this formalism 
there is a doubling of degrees of freedom \cite{Lebellac}. Fields are labeled as ``1'' or ``2''. Fields ``1'' (``2'')  only interact with fields ``1'' (``2'') according to (minus) the original interaction Lagrangian. Fields ``1'' may be converted to fields ``2'', and vice versa, through propagation so that propagators become $2\times 2$ matrices. For instance, for a free scalar field we have
\bea
\Delta (K)=\left (\begin{array}{cc} \frac{1}{K^2-m^2+i\epsilon} & 0\\
                              0 & \frac{-1}{K^2-m^2-i\epsilon}\\
            \end{array} \right )
-2\pi i\, \delta (K^2-m^2)\> \left (\begin{array}{cc}
n_B(k_0) & \theta (-k_0)+n_B(k_0)\\
\theta (k_0)+n_B(k_0) & n_B(k_0) \\ \end{array} \right ),
\label{2.23}
\eea
where $i\Delta (K)$ is the Feynman propagator, and $ n_B(k_0)$ the Bose distribution function.

For the tree-level propagator of the transverse electromagnetic field in the Coulomb gauge, $i \Delta_{ij}(K)$, we have
\begin{equation}
\Delta_{ij}(K)=\left(\delta_{ij}-\frac{k_ik_j}{|{\bf k}|^2}\right)\Delta (K)
\end{equation}
The tree-level $A_0$-propagator matrix in the Coulomb gauge is diagonal, traceless and the ``11'' component coincides with the propagator at zero temperature.

For the tree-level propagator of a Dirac fermion field, $i S(K)$, we have $S(K)\equiv (K\sla +m)\> \tilde \Delta(K)$,
where $\tilde \Delta (K)$ follows from $\Delta (K)$ by replacing
$n_B$ by $-n_F$, $n_F$ being the Fermi-Dirac distribution.

Since we are always in the case $m_\mu>>T$ all vertices involving muons or protons will be of type "1". 
However, vertices involving photons and electrons can be both type "1" and "2". At the order we are calculating, it turns out that
 we only need propagators of the type "11" for the photons, either at tree level or including one-loop self-energies.

For computations that require loop corrected propagators (for example fig. \ref{pNRQCDHTL} and fig. \ref{pNRQEDoneloop}) it is convenient to use the so called Keldysh representation \cite{Thoma:2000dc}. In this representation the retarded, advanced and symmetric propagators are defined as,
\begin{eqnarray}
&\Delta_R=\Delta_{11}-\Delta_{12},\label{R}\\
&\Delta_A=\Delta_{11}-\Delta_{21},\\
&\Delta_S=\Delta_{11}+\Delta_{22},
\end{eqnarray}
Notice from (\ref{2.23}) that at tree level only $\Delta_S$ depends on the temperature.

In order to calculate loop corrections to the $\Delta_{11}$ propagator in an efficient way, we use the following method \cite{Thoma:2000dc}.
\begin{itemize}
\item We compute the $\Delta_R$ propagator, using the fact that for this propagator the Dyson equation is of the zero temperature type (this is not so for $\Delta_S$).
\begin{equation}
\Delta_R=\Delta^0_R+\Delta^0_R\Pi_R\Delta_R,
\end{equation}
where the self-energy $\Pi_R$ is
\begin{equation}
\Pi_R=\Pi_{11}+\Pi_{12},
\end{equation}
and $\Delta^0_R$ is the tree-level retarded propagator obtained according to formulas (\ref{2.23})-(\ref{R}).
\item The advanced propagator $\Delta_A$ is the complex conjugate of the retarded one and the symmetric propagator $\Delta_S$ reads, in the bosonic case,
\begin{equation}
\label{relS}
\Delta_S(K)=[1+2n_B(|k_0|)]sgn(k_0)[\Delta_R(K)-\Delta_A(K)].
\end{equation}
\item Finally, one obtains
\begin{equation}
\label{def11}
\Delta_{11}=\frac{1}{2}(\Delta_R+\Delta_A+\Delta_S).
\end{equation}
\end{itemize}
Note that for $|k_0|\ll T$ the size of the symmetric propagator for bosons is larger than what one would expect from naive power counting
\begin{equation}
\label{BEe}
\Delta_S(K)=\frac{T}{k_0}[\Delta_R(K)-\Delta_A(K)]
\end{equation}
plus terms suppressed by $1/T$. This effect is called Bose-enhancement, and it complicates
the power counting of the EFTs at scales lower than the temperature.

\section{Calculations in Section II.A}
\label{IIA}

All these computations have in common that the starting point is pNRQED. In this effective theory, and in all the theories that are derived from it by integrating out further degrees of freedom, the leading correction to the Hamiltonian is \cite{Pineda:1997bj,Brambilla:2008cx}
\begin{equation}
\label{apepot}
\delta H=ie^2r^i\int\frac{\,d^Dk}{(2\pi)^D}\frac{1}{E-H-k_0+i\epsilon}(k_0^2\Delta_{ij}+k_ik_j\Delta_{00})r^j,
\end{equation}
with $\Delta_{ij}$ and $\Delta_{00}$ being the "11" transverse and $A_0$ photon propagators respectively. $\delta H$ is a potential $\delta V$ (energy-independent) only if $E$ is much smaller than the scales inside the $\Delta$ propagators. Depending on the concrete calculation that one is doing the explicit expression for $\Delta$ may be different.

\subsection{Integrating out the $T$ scale}
\label{apetint}
After integrating out the scale $T$ we will reach $pNRQED_{>T}$. At leading order the difference between $pNRQED$ and $pNRQED_{>T}$ will be a correction on the potential of the type of (\ref{apepot}) where the internal momentum $K$ is of order $T$. 

The LO correction (\ref{vlo}) is obtained using the tree level photon propagator in (\ref{apepot}). If one uses the Coulomb gauge thermal effects only appear in transverse photons. This computation is done in detail for different relations between $T$ and $E$ in \cite{Escobedo:2008sy}.
\begin{equation}
\delta V^{LO}=e^2r^i\int\frac{\,d^Dk}{(2\pi)^D}\frac{k_0^2}{E-H-k_0+i\epsilon}\left(\delta_{ij}-\frac{k_ik_j}{|{\bf k}|^2}\right)2\pi\delta(K^2)n_B(k_0)r^j
\end{equation}

The NLO potential comes from including one-loop corrections to the photon propagator (this corresponds to the diagram of fig. \ref{pNRQCDHTL}). This can be done analytically for $T>>E$. The required expression for the photon propagator are found in section III.B of \cite{Brambilla:2008cx}. Since $T>>E$ the muonic hydrogen propagator can be expanded
\begin{equation}
\frac{1}{E-H-k_0+i\epsilon}\to\frac{1}{-k_0+i\epsilon}-\frac{(E-H)}{(-k_0+i\epsilon)^2}+...\label{E<T}
\end{equation}
Let us compute the first term in this expansion
\begin{equation}
\delta V^{NLOa}=ie^2r^i\int\frac{\,d^Dk}{(2\pi)^D}\frac{1}{-k_0+i\epsilon}(k_0^2\Delta_{ij}^{1-loop}+k_ik_j\Delta_{00}^{1-loop})r^j
\end{equation}
Looking at the expressions of \cite{Brambilla:2008cx} and using the relation (\ref{def11}) one sees that both $\Delta_{ij}$ and $\Delta_{00}$ are even functions in $k_0$, so one can use for the first term of the expansion
\begin{equation}
\frac{i}{-k_0+i\epsilon}\to\pi\delta(-k_0).
\end{equation}
So we only need the photon propagator in the limit $k_0\to 0$ to get the leading order. 

In \cite{Brambilla:2008cx} the propagators is given as an integral of a parameter $q_0$. The best strategy to perform the calculation is to integrate first over internal momentum and leave the integration of this parameter to the end. This computation was done in section IV.B.1 of \cite{Brambilla:2008cx}, here we take the Abelian limit making $C_F=1$ and $C_A=0$.
\begin{equation}
\delta V^{NLOa}=\frac{3}{2}\zeta(3)\frac{\alpha}{\pi}r^2Tm_D^2+i\frac{\alpha}{6}r^2Tm_D^2\left(\frac{1}{\epsilon}+\gamma_E+\log\pi-\log\frac{T^2}{\mu^2}+\frac{2}{3}-4\log 2-2\frac{\zeta^\prime(2)}{\zeta(2)}\right)
\end{equation}

For the next-to-leading order in (\ref{vnlo}), namely the contributions coming from second term in (\ref{E<T}), 
we will restrict ourself to the extraction of the leading IR logarithmic behavior. We use the fact that in the infrared the photon self-energy approaches the HTL limit, so instead, we substitute the complete self-energy $\Pi_{HTL}e^{-\beta^2 k^2}$. The factor $e^{-\beta^2 k^2}$ is introduced to regulate UV divergences and does not affect the leading IR behavior we are interested in. The HTL self-energy have the property that can be written as $\Pi_{HTL}=m_D^2f(k_0/k)$ with $f(x)$ a non trivial function. Performing the change of variables $k_0=kx$ the infrared behavior can be easily extracted. First we compute the contribution from the retarded part of the longitudinal photon propagator.
\begin{equation}
\delta V^{NLOb}_L=ie^2r^i(E-H)\int\frac{\,d^Dk}{(2\pi)^D}\frac{k_ik_j}{k^4}\frac{\Pi^L_R(k_0,k)}{(k_0-i\epsilon)^2}r^j=-\frac{e^2}{2m_\mu}\int\frac{\,d^Dk}{(2\pi)^D}\frac{\Pi^L_R(k_0,k)}{k^2(k_0-i\epsilon)^2}
\end{equation}
We have used that $r^i(E-H)r_i=-\frac{1}{2m_\mu}$ plus terms that vanish on the physical state.
As we are only interested in the leading logarithm behavior, the change $\Pi^L_R(k_0,k)\to\Pi^L_{HTL,R}(k_0,k)e^{-\beta^2 k^2}=m_D^2f^L_R(k_0/k)e^{-\beta^2 k^2}$ can be made, where
\begin{equation}
\label{fRL}
f^L_R(x)=1-\frac{x}{2}\log\left(\frac{x+1+i\epsilon}{x-1+i\epsilon}\right)
\end{equation}
\begin{equation}
\delta V^{NLOb}_L=-\frac{ie^2m_D^2}{2m_\mu}\int\frac{\,d^Dk}{(2\pi)^D}\frac{f^L_R(k_0/k)e^{-\beta^2 k^2}}{k^2(k_0-i\epsilon)^2}
\end{equation}
The equality above is only true as far as the leading IR behavior is concerned. Now with the change $k_0=kx$
\begin{equation}
\delta V^{NLOb}_L=-\frac{ie^2m_D^2}{2m_\mu}\frac{\Omega_{D-1}}{(2\pi)^D}\int_0^\infty\,dkk^{D-5}e^{-\beta^2 k^2}\int_{-\infty}^\infty\frac{\,dxf^L_R(x)}{(x-i\epsilon)^2}=-\frac{ie^2m_D^2}{4m_\mu}\frac{\Omega_{D-1}}{(2\pi)^D}\left(\frac{T}{\mu}\right)^{D-4}\Gamma({D-4\over 2})\int_{-\infty}^\infty\frac{\,dxf^L_R(x)}{(x-i\epsilon)^2}
\label{lIR}
\end{equation}
$\Omega_{D-1}$ is the $D-1$ dimensional solid angle.
Note that the contribution from the advanced part of the longitudinal photon propagator is obtained by replacing $f^L_R(x)\to (f^L_R)^\ast (x)$ in (\ref{lIR}). Since it has all the singularities in the upper half complex plain the corresponding integral gives zero.
The integral over $x$ in (\ref{lIR}) can be done using standard techniques from integration in the complex plane, and the result is 
\begin{equation}
\int_{-\infty}^\infty\frac{\,dx f^L_R(x)}{(x-i\epsilon)^2}=-\pi^2
\label{pi2}
\end{equation}
The contribution from the symmetric part of the longitudinal photon propagator does not produce IR divergences in dimensional regularization. So we do not need to compute it here as we are only interested in the logarithms.

Next we proceed analogously for the transverse photon propagator. As in the longitudinal part, only the retarded plus advanced contribution is infrared sensible. We approximate $\Pi^T_R(k_0,k)=m_D^2f^T_R(k_0/k)e^{-\beta^2 k^2}$ where now,
\begin{equation}
f^T_R(x)=\frac{1}{2}\left[x^2-(x^2-1)\frac{x}{2}\log\left(\frac{x+1+i\epsilon}{x-1+i\epsilon}\right)\right]\; ,
\end{equation}
then
\begin{equation}
\delta V^{NLOb}_T=-\frac{ie^2r^i(E-H)r^j}{2}\int\frac{\,d^Dk}{(2\pi)^D}m_D^2e^{-\beta^2 k^2}\left(\delta_{ij}-\frac{k_ik_j}{k^2}\right)\left[\frac{f^T_R(k_0/k)}{(k_0^2-k^2+isgn(k_0)\epsilon)^2}+\frac{f^T_A(k_0/k)}{(k_0^2-k^2-isgn(k_0)\epsilon)^2}\right]\; .
\end{equation}
The first (second) term in the square brackets have all the singularities in lower (upper) complex $k_0$ half plane, and hence the whole expression vanishes. The contribution from the symmetric piece of the propagator also vanishes for the same reason. Hence,
\begin{equation}
\delta V^{NLO}=0\label{intx0}\; .
\end{equation}

\subsection{The $E$ scale for $T>>E>>m_D$}
\label{TEmD}
In order to obtain (\ref{Egt2}) and (\ref{DecaygT2}) the starting point is $pNRQED_T$. In this effective theory the atom self-energy gives
the correction to the Hamiltonian (\ref{apepot}). Recall that the photon propagators in (\ref{apepot}) must be taken in the HTL approximation.
We use the method of the integration
by regions \cite{Beneke:1997zp,Smirnov:2002pj} in order to evaluate it.
In this integral there are three relevant regions: $k,k_0\sim E$ with $\lambda=k_0-k \sim E$ (we call it off-shell region), $k,k_0\sim E$ with $\lambda\sim m_D$ (collinear region), and $k,k_0\sim m_D$ ($m_D$-region).
 By power counting one can see that the $m_D$-region will be just a higher order correction, so we focus in the regions $k,k_0\sim E$.
 In the off-shell region the atom propagator can not be expanded, but the HTL photon propagator can. For example, the longitudinal photon retarded propagator can be written as
 \begin{equation}
 \Delta_{00,R}=\frac{1}{k^2+m_D^2 f^L_R(k_0/k)},
 \end{equation}
 and, after the compulsory expansion
 \begin{equation}
 \Delta_{00,R}=\frac{1}{k^2}-\frac{m_D^2 f^L_R(k_0/k)}{k^4}+...
 \end{equation}
 The fact that the non-trivial functions appears only in the numerator after the expansion is crucial in order to be able to do the integration analytically. The collinear region does not contribute in the part that is related with longitudinal photons.

The transverse photon retarded propagator is
\begin{equation}
\Delta_{ij,R}=\frac{\left(\delta_{ij}-\frac{k_ik_j}{k^2}\right)}{k_0^2-k^2-m_D^2f^T_R(k_0/k)+i\epsilon}
\end{equation}
In the off-shell region a expansion like the one for the longitudinal photon propagator has to be made. In the collinear region (that has $k,k_0 \sim E$ but $k_0^2-k^2\sim m_D^2$), the atom propagator can not be expanded also, but the HTL photon propagator has to be expanded around the region $|k_0/k|\sim 1$. As in the previous case, this expansion makes it possible to proceed analytically.

 Details of this computation can be found in \cite{MilaT2},

\subsection{Integrating out $m_D$ for $m_D>>E$}
\label{D>mE}
After integrating out $m_D$ we will arrive to what we will call $pNRQED_{>m_D}$. In the photon sector we will have a non trivial action but we will not need it at the level of precision we are working. In the atom sector there appears a correction of the potential in the matching between $pNRQED_{T}$ (or $pNRQED_{>T}$) and $pNRQED_{>m_D}$ of the form of (\ref{apepot}) where the internal momentum is of order $k\sim m_D$.

Because $m_D>>E$ we can also put $E-H=0$ at leading order in the atom propagator as in section \ref{apetint}. Hence, we will need the HTL photon propagator 
in the $k_0\to 0$ limit(i.e. $T>>k>>k_0$). The HTL photon propagators are very well known and can be found in \cite{Lebellac,Thoma:2000dc,Brambilla:2008cx}. In our case the fact that $k_0\to0$ makes
\begin{equation}
k_0^2\Delta_{ij}(K)=0,
\end{equation}
\begin{equation}
k_ik_j\Delta_{00}(K)=k_ik_j\left(\frac{1}{k^2+m_D^2}-\frac{i\pi Tm_D^2}{k(k^2+m_D^2)^2}\right).
\end{equation}
Using this in (\ref{apepot}) the results (\ref{EgT}) and (\ref{DecaygT}) are obtained.

For the next-to-leading order (\ref{DecaygTNLO}) the calculation is very similar to the one we have carried out for the $T$ scale. Like in that case, we will focus on the logarithmic behavior, now in the UV. As an example, we study again the retarded part of the longitudinal photon propagator (note that consistency with eq. (\ref{intx0}) ensures that there will not be a logarithmic contribution from transverse photons)
\begin{equation}
\delta V^{NLOm_D}_L=-ie^2r^i\int\frac{\,d^Dk}{(2\pi)^D}\frac{k_ik_j(E-H)}{(k_0-i\epsilon)^2}\frac{1}{k^2+m_D^2f^L_R(k_0/k)}r^j=\frac{ie^2}{2m_\mu}\int\frac{\,d^Dk}{(2\pi)^D}\frac{k^2}{(k_0-i\epsilon)^2}\frac{1}{k^2+m_D^2f^L_R(k_0/k)}
\end{equation}
using that scale-less integrals in dimensional regularization are zero
\begin{equation}
\delta V^{NLOm_D}_L=-\frac{ie^2m_D^2}{2m_\mu}\int\frac{\,d^Dk}{(2\pi)^D}\frac{f^L_R(k_0/k)}{(k_0-i\epsilon)^2}\frac{1}{k^2+m_D^2f^L_R(k_0/k)}
\end{equation}
Now is when the change $k_0=kx$ becomes useful
\begin{eqnarray}
&\delta V^{NLOm_D}_L=-\frac{ie^2m_D^2}{2m_\mu}\frac{\Omega_{D-1}}{(2\pi)^D}\int_{-\infty}^\infty\frac{\,dxf^L_R(x)}{(x-i\epsilon)^2}\int_0^\infty\frac{\,dkk^{D-3}}{k^2+m_D^2f^L_R(x)}= \\
&=\frac{ie^2m_D^2}{4m_\mu}\frac{\Omega_{D-1}}{(2\pi)^D}\pi Csc\left(\frac{D\pi}{2}\right)\left(\frac{m_D}{\mu}\right)^{D-4}\int_{-\infty}^\infty\frac{\,dx(f^L_R(x))^{D-3}}{(x-i\epsilon)^2} \nonumber
\end{eqnarray}
If one is only interested in the logarithmic behavior, one can put $D-3=1$ in the $x$ integration and use (\ref{pi2}). Formula (\ref{DecaygTNLO}) is readily obtained from the expression above.
 
\subsection{UV behavior for $E\sim m_D$}
Although in this situation we have not been able to obtain an analytic result for (\ref{apepot}), its UV behavior can be easily isolated 
as follows.
\begin{equation}
\frac{1}{E-H-k_0+i\epsilon}=\left(\frac{1}{E-H-k_0+i\epsilon}-\frac{1}{-k_0+i\epsilon}+\frac{(E-H)}{(-k_0+i\epsilon)^2}\right)+\frac{1}{-k_0+i\epsilon}-\frac{(E-H)}{(-k_0+i\epsilon)^2}.
\end{equation}
The piece in brackets
in the right hand side of the equation above leads to a 
ultraviolet finite expression when substituted in (\ref{apepot}). So the ultraviolet divergences arise from the remaining terms in the right hand side
 of this equation. In fact the computation for these UV divergences is exactly the same as in the case $m_D>>E$, which we have carried in section 
\ref{D>mE}.

\section{Calculations in Section III.A}
\label{IIIA}
\subsection{Correction to the Coulomb potential in $pNRQED_T$}
In this part we deal with the matching procedure that has to be done for $T\sim p$. In perturbation theory the $pNRQED$ potential is related to 
the Fourier transform of the longitudinal photon propagator in the limit where $p_0\to 0$ (for $P$ the external momentum of the propagator). 
For this temperature range the propagator, which can be obtained by the procedure outlined in Appendix \ref{review}, is needed for 
$T\sim m_e\sim p>>p_0$. The retarded self-energy reads
\begin{eqnarray}
&\Pi_R(P)=-\frac{2e^2}{\pi^2}\int_0^\infty\frac{\,dkk^2}{\sqrt{k^2+m_e^2}(e^{\beta\sqrt{k^2+m_e^2}}+1)}+\frac{e^2}{\pi^2p}\int_0^\infty\frac{\,dkk(2k^2+2m_e^2-p^2/2)}{\sqrt{k^2+m_e^2}(e^{\beta\sqrt{k^2+m_e^2}}+1)}\log\left(\frac{|-p+2k|}{|p+2k|}\right)- \\ 
&-\frac{2ie^2p_0}{\pi p}\int_{p/2}^\infty\frac{\,dk k}{e^{\beta\sqrt{k^2+m_e^2}}+1}-\frac{ie^2p_0m^2}{\pi p}\frac{1}{e^{\beta\sqrt{p^2/4+m_e^2}}+1}. \nonumber
\end{eqnarray}
For $m_e\to 0$, this self-energy coincides with the Abelian limit of the one found in \cite{Brambilla:2008cx}.

In order to obtain the potential the first step is to use formulas (\ref{def11}) and (\ref{relS}) to get the corrections to the "11" propagator. 
Then the propagator is related with to potential plus the self-energy by the following formula
\begin{equation}
\label{defV}
V(r)=-e^2\int\frac{\,d^{D-1}p}{(2\pi)^{D-1}}(e^{i{\bf p}{\bf r}}-1)\Delta_{11}(p_0=0,p)
\end{equation}
at leading order this gives the Coulomb potential $V(r)=-\frac{\alpha}{r}$. For simplicity we define
\begin{equation}
\label{defVr}
V_r(r)=-e^2\int\frac{\,d^{D-1}p}{(2\pi)^{D-1}}e^{i{\bf p}{\bf r}}\Delta_{11}(p_0=0,p)
\end{equation}
and
\begin{equation}
\label{defVm}
V_m=-e^2\int\frac{\,d^{D-1}p}{(2\pi)^{D-1}}\Delta_{11}(p_0=0,p) 
\end{equation}
such that
\begin{equation}
V(r)=V_r(r)-V_m
\end{equation}
For  the next to leading order we need the corrections to the propagator
\begin{equation}
\delta \Delta_{11}=\Delta_1+\Delta_2+\Delta_3+\Delta_4+\Delta_5
\end{equation}
where
\begin{equation}
\label{defD1}
\Delta_1=-\frac{2e^2}{\pi^2p^4}\int_0^\infty\frac{\,dkk^2}{\sqrt{k^2+m_e^2}(e^{\beta\sqrt{k^2+m_e^2}}+1)}
\end{equation}

\begin{equation}
\label{defD2}
\Delta_2=\frac{2e^2}{\pi^2p^5}\int_0^\infty\frac{\,dkk\sqrt{k^2+m_e^2}}{e^{\beta\sqrt{k^2+m_e^2}}+1}\log\frac{|-p+2k|}{|p+2k|}=
-\frac{2e^2}{\pi^2p^5}\int_0^\infty\frac{\,dkk^2\sqrt{k^2+m_e^2}}{e^{\beta\sqrt{k^2+m_e^2}}+1}\int_{-1}^1\,d\lambda\left(\frac{1}{p-2k\lambda+i\epsilon}+\frac{1}{p-2k\lambda-i\epsilon}\right)
\end{equation}

\begin{eqnarray}
&\Delta_3=-\frac{e^2}{2\pi^2p^3}\int_0^\infty\frac{\,dkk}{\sqrt{k^2+m_e^2}(e^{\beta\sqrt{k^2+m_e^2}}+1)}\log\frac{|-p+2k|}{|p+2k|}= \nonumber \\
&=\frac{e^2}{2\pi^2p^3}\int_0^\infty\frac{\,dkk^2}{\sqrt{k^2+m_e^2}(e^{\beta\sqrt{k^2+m_e^2}}+1)}\int_{-1}^1\,d\lambda\left(\frac{1}{p-2k\lambda+i\epsilon}+\frac{1}{p-2k\lambda-i\epsilon}\right)
\end{eqnarray}

\begin{equation}
\Delta_4=-\frac{4iTe^2}{\pi p^5}\int_{p/2}^\infty\frac{\,dkk}{e^{\beta\sqrt{k^2+m_e^2}}+1}
\end{equation}

\begin{equation}
\Delta_5=-\frac{2iTe^2m_e^2}{\pi p^5}\frac{1}{e^{\beta\sqrt{p^2/4+m_e^2}}+1}
\end{equation}
The contribution of (\ref{defD1}) to (\ref{defV}) leads to the first term in the first line of (\ref{potmno0}).
In order to calculate the contribution (\ref{defD2}) to (\ref{defV}),
it is convenient to leave the integration over the internal momentum $k$ to the end. Consider then,
\begin{equation}
\label{step1}
-\frac{e^2}{2}\int_{-1}^1\,d\lambda\int\frac{\,d^{D-1}p}{(2\pi)^{D-1}}\frac{e^{i{\bf p}{\bf r}}}{p^5}\left[\frac{1}{p-2k\lambda+i\epsilon}+\frac{1}{p-2k\lambda-i\epsilon}\right]
\end{equation}
and use
\begin{equation}
\label{expnoir}
\frac{1}{p-2k\lambda+i\epsilon}=-\frac{1}{2k\lambda-i\epsilon}-\frac{p}{(2k\lambda-i\epsilon)^2}-\frac{p^2}{(2k\lambda-i\epsilon)^3}-\frac{p^3}{(2k\lambda-i\epsilon)^4}+\frac{p^4}{(2k\lambda-i\epsilon)^4(p-2k\lambda+i\epsilon)}
\end{equation}
in (\ref{step1}), 
\begin{eqnarray}
&\frac{e^2}{2}\int_{-1}^1\,d\lambda\left(\frac{1}{(2k\lambda-i\epsilon)^2}+\frac{1}{(2k\lambda+i\epsilon)^2}\right)\int\frac{\,d^{D-1}p}{(2\pi)^{D-1}}\frac{e^{i{\bf p}{\bf r}}}{p^4}+\frac{e^2}{2}\int_{-1}^1\,d\lambda\left(\frac{1}{(2k\lambda-i\epsilon)^4}+\frac{1}{(2k\lambda+i\epsilon)^4}\right)\int\frac{\,d^{D-1}p}{(2\pi)^{D-1}}\frac{e^{i{\bf p}{\bf r}}}{p^2}-... \nonumber \\
&...-\frac{e^2}{2}\int_{-1}^1\,d\lambda\int\frac{\,d^{D-1}p}{(2\pi)^{D-1}}\frac{e^{i{\bf p}{\bf r}}}{p}\left(\frac{1}{(2k\lambda-i\epsilon)^4(p-2k\lambda+i\epsilon)}+\frac{1}{(2k\lambda+i\epsilon)^4(p-2k\lambda-i\epsilon)}\right)
\end{eqnarray}
Some terms vanish because of the symmetry $\lambda \to -\lambda$. The integral over $p$ in the first and the second term is straight-forward. 
By using the symmetries in the $\lambda$ and $p$ variables, the third term can be simplified as follows,
\begin{eqnarray}
&\frac{ie^2}{32\pi^2r}\int_{-1}^1\,d\lambda\int_{-\infty}^\infty\,dp(e^{ipr}-e^{-ipr})\left(\frac{1}{(2k\lambda-i\epsilon)^4(p-2k\lambda+i\epsilon)}+\frac{1}{(2k\lambda+i\epsilon)^4(p+2k\lambda+i\epsilon)}+...\right. \nonumber \\
&\left. ...+\frac{1}{(2k\lambda+i\epsilon)^4(p-2k\lambda-i\epsilon)}+\frac{1}{(2k\lambda-i\epsilon)^4(p+2k\lambda-i\epsilon)}\right)\; .
\end{eqnarray}
At this point the integral over $p$ can be done using standard techniques of complex analysis,
\begin{eqnarray}
&-\frac{e^2}{8\pi r}\int_{-1}^1\,d\lambda\left[\frac{e^{i2k\lambda}}{(2k\lambda+i\epsilon)^4}+\frac{e^{-i2k\lambda}}{(2k\lambda-i\epsilon)^4}\right]= \nonumber \\
&=-\frac{e^2}{8\pi r}\int_{-1}^1\,d\lambda\cos(2k\lambda)\left(\frac{1}{(2k\lambda+i\epsilon)^4}+\frac{1}{(2k\lambda-i\epsilon)^4}\right)-\frac{ie^2}{8\pi r}\int_{-1}^1\,d\lambda\sin(2k\lambda)\left(\frac{1}{(2k\lambda+i\epsilon)^4}-\frac{1}{(2k\lambda-i\epsilon)^4}\right)
\end{eqnarray}
Hence, our final result for (\ref{step1}) reads, 
\begin{equation}
-\frac{e^2}{8\pi r}\int_{-1}^1\,d\lambda\left(\frac{1}{(2k\lambda+i\epsilon)^4}+\frac{1}{(2k\lambda-i\epsilon)^4}\right)(\cos(2k\lambda r)-1+2k^2\lambda^2r^2)-\frac{ie^2}{8\pi r}\int_{-1}^1\,d\lambda\sin(2k\lambda r)\left(\frac{1}{(2k\lambda+i\epsilon)^4}-\frac{1}{(2k\lambda-i\epsilon)^4}\right)
\end{equation}
After performing the integration in $\lambda$ we obtain the second and third line in (\ref{potmno0}). 
$\Delta_3$ can be computed in a very similar way, 
and leads to the second term of the first line in (\ref{potmno0}).
Let us next consider $\Delta_4$,
\begin{eqnarray}
&\frac{4iTe^4}{\pi}\int\frac{\,d^{D-1}p}{(2\pi)^{D-1}}\frac{e^{i{\bf p}{\bf r}}}{p^5}\int_{p/2}^\infty\frac{\,dkk}{e^{\beta\sqrt{k^2+m_e^2}}+1}= \nonumber \\
&=\frac{4iTe^4}{\pi}\int\frac{\,d^{D-1}p}{(2\pi)^{D-1}}\frac{e^{i{\bf p}{\bf r}}}{p^5}\left[\int_0^\infty\frac{\,dkk}{e^{\beta\sqrt{k^2+m_e^2}}+1}-\frac{p^2}{8(e^{\beta m_e}+1)}+\left(\int_{p/2}^\infty\frac{\,dkk}{e^{\beta\sqrt{k^2+m_e^2}}+1}-\int_0^\infty\frac{\,dkk}{e^{\beta\sqrt{k^2+m_e^2}}+1}+\frac{p^2}{8(e^{\beta m_e}+1)}\right)\right]\; .
\end{eqnarray}
We have separated above the pieces that lead to infrared divergences from the ones that do not. The first term inside the square brackets gives 
the fourth line in (\ref{potmno0}), and the second term  together with $V_m$ gives the fifth line of (\ref{potmno0}). The rest of terms in
 the square brackets give the sixth and seventh line of (\ref{potmno0}).
$\Delta_5$ can be computed in a very similar way, and gives the remaining terms of (\ref{potmno0}).

\section{Computation of the HTL retarded self-energy for the longitudinal photon in the $m_e\ne 0$ case}
\label{HTLme}
There are some subtle points in the computation that leads to (\ref{DecaymgT}), 
which do not arise in the massless case and are worth elaborating upon.
Let us start from the formula (37) of \cite{Thoma:2000dc} for the massive case, 
\begin{equation}
\Pi_R^L(P)=-2ie^2\int\frac{d^Dk}{(2\pi)^D}(q_0k_0+{\bf q}{\bf k}+m_e^2)[\tilde{\Delta}_S(Q)\tilde{\Delta}_R(K)+\tilde{\Delta}_A(Q)\tilde{\Delta}_S(K)]
\end{equation}
$Q=P-K$. It is customary to make the change $K \to -Q$ in the second term to get a simplified expression that reduces to twice the first term.
However, the terms proportional to $m_e^2$, which do not exist in the massless case, have a stronger IR sensibility than the remaining ones. This leads 
to ill-defined expressions in the HTL approximation for $T\sim m_e$. These expressions must be properly defined in order to get consistent results 
before and after the shift $K \to -Q$ has been carried out.

To see this in detail we work out just a small part of the computation which illustrates the point, namely the part of $\Im \Pi_R^L(P)$ that comes 
only from the $m_e^2$ term in the numerator. We call this term $ \Pi_{m_e^2}(P)$,
\begin{equation}
\label{pime2}
\Pi_{m_e^2}(P)=-2ie^2m_e^2\int\frac{\,d^Dk}{(2\pi)^D}[\tilde{\Delta}_S(Q)\tilde{\Delta}_R(K)+\tilde{\Delta}_A(Q)\tilde{\Delta}_S(K)]
\; .
\end{equation}
One can use the shift $K \to -Q$ to get a simplified expression,
\begin{equation}
\Pi_{m_e^2}(P)=-4ie^2m_e^2\int\frac{\,d^Dk}{(2\pi)^D}\tilde{\Delta}_A(Q)\tilde{\Delta}_S(K)\; .
\end{equation}
By carrying out the HTL expansion and taking into account that we must expand also $p_0$ ( $p\gg p_0$ in the computation of the potential), we obtain,
\begin{equation}
\Pi_{m_e^2}(P)=8e^2m_e^2p^2\int\frac{\,d^{D-1}k}{(2\pi)^{D-1}}\frac{n_F(\sqrt{k^2+m_e^2})}{\sqrt{k^2+m_e^2}}\left[\frac{1}{(2{\bf k}{\bf p}-i\epsilon)^2}+\frac{4p_0\sqrt{k^2+m_e^2}}{(2{\bf k}{\bf p}-i\epsilon)^3}+...\right]
\; .
\end{equation}
Notice that the last term is ill-defined in the IR (this is apparent if spherical coordinates are used). Let us focus on the imaginary part,
\begin{equation}
\Im \Pi_{m_e^2}=-16ie^2m_e^2p_0p^2\int\frac{\,d^{D-1}k}{(2\pi)^{D-1}}n_F(\sqrt{k^2+m_e^2})\left[\frac{1}{(2{\bf k}{\bf p}-i\epsilon)^3}-\frac{1}{(2{\bf k}{\bf p}+i\epsilon)^3}\right]\; ,
\end{equation}
For illustration purposes, we will make this computation in two ways
\begin{itemize}
\item Using $\,d^{D-1}k=\,d\Omega_{D-2}\,d(\cos\theta)k^{D-2}\,dk$, where
$\theta$ is the angle between the internal momentum and the external one.
\begin{equation}
\Im\Pi_{m_e^2}=-32ie^2m_e^2p_0p^2\frac{\Omega_{D-2}}{(2\pi)^{D-1}}\int_0^\infty\,dkk^{D-2}n_F(\sqrt{k^2+m_e^2})\int_0^\pi\frac{\,d(\cos\theta)}{(2kp\cos\theta-i\epsilon)^3}
\; ,
\end{equation}
and performing the angular integration one arrives at
\begin{equation}
\Im\Pi_{m_e^2}=4\pi e^2\frac{m_e^2p_0}{p}\frac{\Omega_{D-2}}{(2\pi)^{D-1}}\int_0^\infty\,dkn_F(\sqrt{k^2+m_e^2})k^{D-4}\delta(k)
\; .
\end{equation}
This expression has an end-point singularity, but one can skip it in dimensional regularization by choosing $D-4>0$. So the result is zero in this way.

\item Using $\,d^{D-1}k=\,d^{D-2}k_\perp\,dk_z$

We choose $z$ to be the direction parallel to the external momenta.
\begin{equation}
\Im\Pi_{m_e^2}=-16ie^2m_e^2p_0p^2\frac{\Omega_{D-2}}{(2\pi)^{D-1}}\int_0^\infty\,dk_\perp k_\perp^{D-3}n_F(\sqrt{k^2_\perp+k^2_z+m_e^2})\int_{-\infty}^\infty\,dk_z\left[\frac{1}{(2pk_z-i\epsilon)^3}-\frac{1}{(2pk_z+i\epsilon)^3}\right]
\end{equation}
Notice that the integrand vanishes for all $k_z$ except when $k_z\sim 0$, so one may substitute 
\begin{equation}
n_F(\sqrt{k_\perp^2+k_z^2+m_e^2})\to n_F(\sqrt{k_\perp^2+m_e^2})+\frac{dn_F}{dE}\frac{k_z^2}{2\sqrt{k_\perp^2+m_e^2}}
\; .
\end{equation}
This simplifies the computation a lot because complex plane integration techniques can be applied,
\begin{equation}
\Im\Pi_{m_e^2}=4\pi e^2m_e^2p_0\frac{\Omega_{D-2}}{(2\pi)^{D-1}}\int_0^\infty\frac{\,dk_\perp k_\perp^{D-3}}{\sqrt{k_\perp^2+m_e^2}}\frac{dn_F}{dE}\int_{-\infty}^\infty\,dk_z\delta(2pk_z)
\end{equation}
This expression does not have any end-point singularity, and in fact it does not need a regularization anymore,
\begin{equation}
\Im\Pi_{m_e^2}=-\frac{e^2m_e^2p_0}{2\pi p(e^{\beta m_e}+1)}
\label{good}
\end{equation}
\end{itemize}
This is indeed the expected result, that eventually leads to terms contributing to the second line of (\ref{DecaymgT}). We arrive then at the 
paradoxical situation in which the final result depends on the precise way DR is implemented, either in spherical coordinates
or in Cartesian ones.
The apparent contradiction is resolved by noticing that DR in spherical coordinates does not allow for the shift $K \to -Q$.
In order to show this is the actual reason for it, let us start now from eq. (\ref{pime2})
\begin{equation}
\Im\Pi_{m_e^2}(P)=-4\pi^2e^2m_e^2\int\frac{\,d^Dk}{(2\pi)^D}\delta(K^2-m_e^2)\delta((K-P)^2-m_e^2)[sgn(k_0-p_0)(1-2n_F(|k_0|))-sgn(k_0)(1-2n_F(|k_0-p_0|))]
\; .
\end{equation}
The integral over $k_0$ is straight-forward,
\begin{eqnarray}
&\Im\Pi_{m_e^2}(P)=-2e^2m_e^2\pi\int\frac{\,d^{D-1}k}{(2\pi)^{D-1}}\frac{1}{\sqrt{k^2+m_e^2}}\left[(n_F(\sqrt{k^2+m_e^2}-p_0)-n_F(\sqrt{k^2+m_e^2}))\delta(p_0^2-p^2-2p_0\sqrt{k^2+m_e^2}+2{\bf k}{\bf p})-... \right. \nonumber \\ 
&\left. ...-(n_F(\sqrt{k^2+m_e^2}+p_0)-n_F(\sqrt{k^2+m_e^2}))\delta(p_0^2-p^2+2p_0\sqrt{k^2+m_e^2}+2{\bf k}{\bf p})\right]
\; .
\end{eqnarray}
So far we have used the complete expression for the self-energy. We apply next HTL expansion and also $p\gg p_0$, like above,
\begin{equation}
\label{ipime}
\Im\Pi_{m_e^2}(P)=-2e^2m_e^2\pi\int\frac{\,d^{D-1}k}{(2\pi)^{D-1}}\frac{2p_0e^{\beta\sqrt{k^2+m_e^2}}}{T\sqrt{k^2+m_e^2}(e^{\beta\sqrt{k^2+m_e^2}}+1)^2}\delta(2{\bf k}{\bf p})
\; .
\end{equation}
From this expression, no matter if one uses spherical or Cartesian coordinates, one arrives at,
\begin{equation}
\Im\Pi_{m_e^2}=-\frac{e^2m_e^2p_0}{2\pi p(e^{\beta m_e}+1)}\; ,
\end{equation}
which agrees with (\ref{good}).
If the expansion for small $p_0$ is not carried out the same result is found with the three methods, because then no regularization is needed.

The conclusion is that when the formula for the retarded self-energy in the massive case \cite{Escobedo:2008sy}, i.e.
\begin{equation}
\Pi_R^L(P)=4e^2\int\frac{\,d^{D-1}k}{(2\pi)^3\sqrt{k^2+m_e^2}}\frac{1}{e^{\beta\sqrt{k^2+m_e^2}}+1}\frac{p^2-\frac{({\bf p}{\bf k})^2}{k^2+m_e^2}}{\left(p_0-\frac{{\bf p}{\bf k}}{\sqrt{k^2+m_e^2}}+i\epsilon\right)^2}
\; ,
\end{equation}
is expanded for $p\gg p_0$, DR must be used in Cartesian coordinates in order to properly regulate IR divergences (i.e to be consistent with the 
shift of momenta carried out at some point in order to get the expression above).  
This point was overlooked when calculating the potential in formula (57) of \cite{Escobedo:2008sy}, and terms analogous to (\ref{ipime}) 
 were missed. The correct formula is obtained by making the following substitution in (57) of \cite{Escobedo:2008sy},
\begin{equation}
g(m\beta)\to g(m\beta)+\frac{m^2\beta^2}{2(e^{\beta m}+1)}
\label{correction}
\end{equation}

\bibliography{./muonic}{}

\begin{thebibliography}{99}















\bibitem{Escobedo:2008sy}
  M.~A.~Escobedo and J.~Soto,
  Phys. Rev. A {\bf 78}, 032520 (2008).

\bibitem{Laine:2006ns}
  M.~Laine, O.~Philipsen, P.~Romatschke and M.~Tassler,
  JHEP {\bf 0703}, 054 (2007)
  [arXiv:hep-ph/0611300].

\bibitem{Laine:2007gj}
  M.~Laine,
  JHEP {\bf 0705}, 028 (2007)
  [arXiv:0704.1720 [hep-ph]].

\bibitem{Burnier:2007qm}
  Y.~Burnier, M.~Laine and M.~Vepsalainen,
  JHEP {\bf 0801}, 043 (2008)
  [arXiv:0711.1743 [hep-ph]].

\bibitem{Beraudo:2007ky}
  A.~Beraudo, J.~P.~Blaizot and C.~Ratti,
  Nucl.\ Phys.\  A {\bf 806}, 312 (2008)
  [arXiv:0712.4394 [nucl-th]].

\bibitem{Brambilla:2008cx}
  N.~Brambilla, J.~Ghiglieri, A.~Vairo and P.~Petreczky,
  Phys.\ Rev.\  D {\bf 78}, 014017 (2008)
  [arXiv:0804.0993 [hep-ph]].

\bibitem{Burnier:2008ia}
  Y.~Burnier, M.~Laine and M.~Vepsalainen,
  JHEP {\bf 0902}, 008 (2009)
  [arXiv:0812.2105 [hep-ph]].



\bibitem{PSI}
  F.~Kottmann et al.,
Hyperf.\ Int.\  {\bf 138} (2001) 55.

\bibitem{DiGiacomo:1969tj}
  A.~Di Giacomo,
  Nucl.\ Phys.\  B {\bf 11} (1969) 411.

\bibitem{PSI2}
  R.~Pohl et al.,
Nature Vol. 466, pag. 213

\bibitem{Ivanov:2009aa}
  V.~G.~Ivanov, E.~Y.~Korzinin and S.~G.~Karshenboim,
  Phys.\ Rev.\ A {\bf 80} (2009) 022510,
  arXiv:0905.4471 [physics.atom-ph].

\bibitem{Borie:2004fv}
  E.~Borie,
Phys.\ Rev.\ A {\bf 71} (2005) 032508,  
arXiv:physics/0410051.

\bibitem{Kinoshita:1998jg}
  T.~Kinoshita and M.~Nio,
  Phys.\ Rev.\  D {\bf 60}, 053008 (1999)
  [arXiv:hep-ph/9812443].

\bibitem{Kinoshita:1998jf}
  T.~Kinoshita and M.~Nio,
  Phys.\ Rev.\ Lett.\  {\bf 82}, 3240 (1999)
  [arXiv:hep-ph/9812442].

\bibitem{Pachucki:1996zza}
  K.~Pachucki,
  Phys.\ Rev.\  A {\bf 53}, 2092 (1996).

\bibitem{Pineda:2002as}
  A.~Pineda,
  Phys.\ Rev.\  C {\bf 67} (2003) 025201
  [arXiv:hep-ph/0210210].
	    
\bibitem{Klines}
  M.~Bregant et al.,
Phys.\ Lett.\ A {\bf 241} (1998) 344.

\bibitem{Fujiwara:2000yf}
  M.~C.~Fujiwara {\it et al.}  [TRIUMF Muonic Hydrogen Collaboration],
  arXiv:nucl-ex/0101007.

\bibitem{Shen:2002zza}
  B.~Shen and J.~Meyer-ter-Vehn,
  Phys.\ Rev.\  E {\bf 65}, 016405 (2002).

\bibitem{Aksenov:2009dy}
  A.~G.~Aksenov, R.~Ruffini and G.~V.~Vereshchagin,
  arXiv:0901.4837 [astro-ph.HE].

\bibitem{Aksenov}
  A.~G.~Aksenov, R.~Ruffini and G.~V.~Vereshchagin,
 Phys.\ Rev.\ Lett.\ {\bf 99}, 125003 (2007).  



\bibitem{Thoma:2008my}
  M.~H.~Thoma,
  arXiv:0801.0956 [physics.plasm-ph].

\bibitem{hollberg}
L.~Hollberg and J.~L.~Hall,
Phys. Rev. Lett. {\bf 53}, 230 (1984).

\bibitem{Blaizot:2001nr}
  J.~P.~Blaizot and E.~Iancu,
  Phys.\ Rept.\  {\bf 359}, 355 (2002)
  [arXiv:hep-ph/0101103].

\bibitem{Rischke:2003mt}
  D.~H.~Rischke,
  Prog.\ Part.\ Nucl.\ Phys.\  {\bf 52}, 197 (2004)
  [arXiv:nucl-th/0305030].

\bibitem{Thoma:2008gh}
  M.~H.~Thoma,
  J.\ Phys.\ A  {\bf 42}, 214004 (2009)
  [arXiv:0809.1507 [hep-ph]].

\bibitem{Caswell:1985ui}
  W.~E.~Caswell and G.~P.~Lepage,
  Phys.\ Lett.\  B {\bf 167}, 437 (1986).

\bibitem{Pineda:1997bj}
  A.~Pineda and J.~Soto,
  Nucl.\ Phys.\ Proc.\ Suppl.\  {\bf 64}, 428 (1998)
  [arXiv:hep-ph/9707481].

\bibitem{Braaten:1991gm}
  E.~Braaten and R.~D.~Pisarski,
  Phys.\ Rev.\  D {\bf 45}, 1827 (1992).

\bibitem{Lebellac}
M.~Le Bellac,
``Thermal Field Theory,''
{\it  Cambridge, UK: Cambridge University Press (1996) 256 p.}

\bibitem{Thoma:2000dc}
  M.~H.~Thoma,
  arXiv:hep-ph/0010164.


\bibitem{Pineda:2002bv}
  A.~Pineda,
  Phys.\ Rev.\  A {\bf 66}, 062108 (2002)
  [arXiv:hep-ph/0204213].

\bibitem{Pineda:1997ie}
  A.~Pineda and J.~Soto,
  Phys.\ Lett.\  B {\bf 420} (1998) 391
  [arXiv:hep-ph/9711292].

\bibitem{Manohar:1997qy}
  A.~V.~Manohar,
  Phys.\ Rev.\  D {\bf 56} (1997) 230
  [arXiv:hep-ph/9701294].

\bibitem{Jentschura}
U.~ D.~ Jentschura, M.~ Haas
Phys.\ Rev.\ A {\bf 78} (2008) 042504


\bibitem{Beneke:1997zp}
  M.~Beneke and V.~A.~Smirnov,
  Nucl.\ Phys.\  B {\bf 522}, 321 (1998)
  [arXiv:hep-ph/9711391].

\bibitem{Smirnov:2002pj}
  V.~A.~Smirnov,
  Springer Tracts Mod.\ Phys.\  {\bf 177}, 1 (2002).

\bibitem{MilaT2}
  N.~Brambilla, M.~A.~Escobedo, J.~Ghiglieri, J.~Soto and A.~Vairo,
  JHEP {\bf 1009} (2010) 038
  [arXiv:1007.4156 [hep-ph]].
  


\bibitem{Bodeker:1998de}
  D.~Bodeker,
  arXiv:hep-ph/9909375.

\bibitem{Laine:2007qy}
  M.~Laine, O.~Philipsen and M.~Tassler,
  JHEP {\bf 0709}, 066 (2007)
  [arXiv:0707.2458 [hep-lat]].

\bibitem{Eiras:2000rh}
  D.~Eiras and J.~Soto,
  Phys.\ Lett.\  B {\bf 491}, 101 (2000)
  [arXiv:hep-ph/0005066].

\bibitem{Pinedathesis}
  A.~Pineda,Phd. Thesis, Universitat de Barcelona (1998).



\end{thebibliography}
\bibliographystyle{h-physrev}

\end{document}